%% file: xKS.tex
\documentclass[12pt]{iopart}

\usepackage{iopams}
\usepackage{cite}

\usepackage{multirow}

\def \half {\frac{1}{2}}
\def \itilde {\tilde{\imath}}
\def \jtilde {\tilde{\jmath}}
\def \d {\rmd}
\def \D {\mathrm{D}}
\def \T {\mbox{\ensuremath{\bigtriangleup}}}
\def \H {\mathcal{H}}
\def \K {\mathcal{K}}
\def \bn {\bi{n}}
\def \bk {\bi{k}}
\def \bl {\bell}
\def \bm {\bi{m}}

\newcommand{\M}[2] {\mbox{\ensuremath{\stackrel{#1}{M}_{#2}}}}

\newcommand{\bM}[2] {\mbox{\ensuremath{\stackrel{#1}{\bar{M}}_{#2}}}}

\newcommand{\pp}{{\it pp\,}-}
\newcommand{\eqref}[1] {\eref{#1}}

\newtheorem{proposition}{Proposition}
\newtheorem{corollary}[proposition]{Corollary}

\begin{document}

\title {Extended Kerr--Schild spacetimes: General properties and some explicit examples}

\author{Tom\'a\v{s} M\'alek$^1$}

\address{$^1$ Institute of Mathematics, Academy of Sciences of the Czech Republic,\\
\v{Z}itn\'a 25, 115 67 Praha 1, Czech Republic}
\ead{malek@math.cas.cz}

\begin{abstract}
We study extended Kerr--Schild (xKS) spacetimes, i.e.\ an extension of the
Kerr--Schild (KS) ansatz where, in addition to the null KS vector, a spacelike
vector field appears in the metric. In contrast to the KS case, we obtain only
a necessary condition under which the KS vector is geodetic. However, it is shown
that this condition becomes sufficient if we appropriately restrict the geometry
of the null KS and spacelike vector. It turns out that xKS spacetimes with
a geodetic KS vector are of Weyl type I and the KS vector has the same optical
properties in the full and background spacetimes. In the case of Kundt xKS
spacetimes, the compatible Weyl types are further restricted and a few examples
of such metrics are provided. The relation of \pp waves to the class of Kundt xKS
spacetimes is briefly discussed. We also show that the CCLP black hole, as
an example of an expanding xKS spacetime, is of Weyl type I$_i$, which specializes
to type D in the uncharged or non-rotating limit, and that its optical matrix
satisfies the optical constraint.
\end{abstract}

%\pacs{???}
%\submitto{\CQG}
%\maketitle

\input{xKS_introduction}

\input{xKS_general_KS_vector}

\input{xKS_geodetic_KS_vector}

\input{xKS_kundt}

\input{xKS_expanding}

\input{xKS_conclusion}

\ack

The author would like to thank to Jan Nov\'ak, Vojt\v{e}ch Pravda and Alena
Pravdov\'a for reading the manuscript and for their helpful comments. This work
was supported by the Albert Einstein Center for gravitation and astrophysics,
Czech Science Foundation 14-37086G.

\appendix

\input{xKS_Ricci_and_Riemann}

\input{xKS_general_kundt}

\section*{References}
\bibliography{bibtex,mypapers}
\bibliographystyle{utcaps}

\end{document}

%% file: xKS_introduction.tex
\section{Introduction}

The Einstein field equations constitute a very complex system of partial
differential equations of the second order and finding exact solutions is
a considerably non-trivial task especially in dimension $n > 4$. One of the
possible approaches to this problem is to assume a convenient form for the
unknown metric in order to simplify subsequent calculations. An important
example representing such a technique is the Kerr--Schild (KS) ansatz
\cite{KerrSchild1965}
\begin{equation}
  g_{ab} = \eta_{ab} - 2 \H k_a k_b,
  \label{KS:KSansatz}
\end{equation}
where $\H$ is a scalar function and $\bk$ is a null vector field with respect
to both the Minkowski background metric $\eta_{ab}$ and the full metric $g_{ab}$.
Despite its simplicity, the class of KS metrics contains many physically
important exact solutions of four-dimensional general relativity, such as
the Kerr black hole, the charged Kerr--Newman black hole, the Vaidya radiating
star, Kinnersley photon rocket, \pp waves and also some of their higher dimensional
analogues (see e.g. \cite{PhD}). In fact, the KS ansatz has led to the discovery
of the rotating black holes in higher dimensions \cite{MyersPerry1986,GibbonsLuPagePope2004}
with vanishing and non-vanishing cosmological constant, respectively. The KS
ansatz has been also successfully applied in the context of higher order gravities
such as the Gauss--Bonnet theory \cite{Anabalonetal2009,AnabalonDeruelleTempoTroncoso2010}
or quadratic gravity \cite{GulluGursesSismanTekin2011,GursesSismanTekin2012}.

Various properties of KS spacetimes with flat background and generalized Kerr--Schild
(GKS) spacetimes with (anti-)de Sitter backgrounds in arbitrary dimension have
been studied in \cite{OrtaggioPravdaPravdova2008} and \cite{MalekPravda2010},
respectively. It has turned out, for instance, that the GKS metric with a geodetic
null congruence $\bk$ leads to algebraically special solutions. The aim of this
paper is to investigate general properties of an extension of the original KS
ansatz referred to as the extended Kerr--Schild (xKS) ansatz that we define as
a metric of the form
\begin{equation}
  g_{ab} = \bar{g}_{ab} - 2 \H k_a k_b - 2 \K k_{(a} m_{b)},
  \label{xKS:metric}
\end{equation}
where $\H$, $\K$ are scalar functions, the background metric $\bar{g}_{ab}$
represents a maximally symmetric vacuum, i.e.\ a Minkowski or (anti-)de Sitter
spacetime, $\bk$ is a null vector and $\bm$ is a spacelike unit vector, both with
respect to the full metric
\begin{equation}
  \fl
  k^a k_a \equiv g_{ab} k^a k^b = 0, \qquad
  k^a m_a \equiv g_{ab} k^a m^b = 0, \qquad
  m^a m_a \equiv g_{ab} m^a m^b = 1.
  \label{xKS:metric:km}
\end{equation}
It immediately follows from the form of the xKS metric \eqref{xKS:metric} that
the same also holds with respect to the background metric
\begin{equation}
  \bar{g}_{ab} k^a k^b = 0, \qquad
  \bar{g}_{ab} k^a m^b = 0, \qquad
  \bar{g}_{ab} m^a m^b = 1
\end{equation}
and that the inverse metric can be simply expressed as
\begin{equation}
  g^{ab} = \bar{g}^{ab} + \left( 2 \H - \K^2 \right) k^a k^b + 2 \K k^{(a} m^{b)}.
  \label{xKS:inversemetric}
\end{equation}

Note that our definition of the xKS ansatz \eqref{xKS:metric} slightly differs
from those ones in \cite{AlievCiftci2008,EttKastor2010}. In our notation, the xKS
metric \eqref{xKS:metric} reduces to the GKS form \cite{MalekPravda2010} for
$\K = 0$, moreover, the spacelike vector $\bm$ is normalized to unity since we
will identify it with one of the frame vectors.

The xKS ansatz has been already studied in \cite{EttKastor2010} through the method
of perturbative expansion. It has turned out that the vacuum Einstein field
equations truncate beyond a certain low order in the expansion around the flat
background metric similarly as in the case of KS spacetimes. In this paper, we
employ the higher dimensional generalization of the Newman--Penrose formalism
\cite{PravdaPravdovaColeyMilson2004,OrtaggioPravdaPravdova2007} and the algebraic
classification of the Weyl tensor in higher dimensions based on the existence of
preferred null directions and their multiplicity
\cite{Coleyetal2004,MilsonColeyPravdaPravdova2004}, see also the recent reviews
\cite{Reall2011,OrtaggioPravdaPravdova2012}. These tools allow us to formulate
some statements about geodeticity and optical properties of the null congruence
$\bk$ and admissible Weyl types of xKS spacetimes. The results of such analysis
could be found helpful in obtaining new exact solutions of the xKS form.

First, let us provide a motivation for studying xKS spacetimes. As will be shown
in section \ref{sec:xKS:geodetic}, one of the reasons why to consider the xKS
ansatz \eqref{xKS:metric} is that such metrics cover more general algebraic types
than GKS metrics. Recall the results of \cite{MalekPravda2010} that GKS spacetimes
with a geodetic Kerr--Schild vector $\bk$ without any further assumptions are of
Weyl type II. Expanding Einstein GKS spacetimes are compatible only with type D
or genuine type II unless conformally flat. Non-expanding Einstein GKS spacetimes
are of type N and belong to the Kundt class. Therefore, expanding Einstein xKS
spacetimes could include black hole solutions of more general Weyl types than II,
for instance black rings \cite{EmparanReall2001} that are of type I$_i$
\cite{PravdaPravdova2005}, and non-expanding Einstein xKS spacetimes do indeed
contain Kundt metrics of more general types than N as will be shown in section
\ref{sec:xKS:Kundt}.

Furthermore, unlike the static charged black hole, rotating charged black hole
as an exact solution of higher dimensional Einstein--Maxwell theory is unknown.
The four-dimensional Kerr--Newman black hole can be cast to the KS form with
the flat background metric $\eta_{ab}$, the function $\H$, and the Kerr--Schild
vector $\bk$ given by \cite{DebneyKerrSchild1969}
\begin{eqnarray}
  \eta_{ab} \, \d x^a \, \d x^b = -\d t^2 + \d x^2 + \d y^2 + \d z^2, \\
  \mathcal{H} = - \frac{r^2}{r^4 + a^2 z^2} \left( Mr - \frac{Q^2}{2} \right), \\
  k_a \, \d x^a = \d t
    + \frac{rx+ay}{r^2 + a^2} \, \d x
    + \frac{ry-ax}{r^2 + a^2} \, \d y
    + \frac{z}{r} \, \d z,
\end{eqnarray}
where $M$, $Q$ are mass and charge of the black hole, respectively. The coordinate
$r$ satisfies
\begin{equation} 
  \frac{x^2+y^2}{r^2 + a^2} + \frac{z^2}{r^2} = 1
\end{equation}
and the vector potential is proportional to the Kerr--Schild vector $\bk$
\begin{equation} 
  A = \frac{Qr^3}{r^4 + a^2z^2} \bk.
\end{equation}
However, the attempt to generalize this solution to higher dimensions using the KS
form of the Myers--Perry black hole has failed \cite{MyersPerry1986}. It has
turned out that a vector potential proportional to $\bk$ cannot simultaneously
satisfy the corresponding Einstein and Maxwell field equations.
It is also known that a straightforward generalization of five-dimensional
rotating black hole solutions of general relativity in the GKS form to the
Gauss--Bonnet theory \cite{Anabalonetal2009} does not represent rotating black
holes \cite{AnabalonDeruelleTempoTroncoso2010} and therefore such spacetimes were
so far studied only numerically \cite{BrihayeRadu2008,BrihayeKleihausKunzRadu2010}
or in the limit of small angular momentum \cite{KimCai2007}.
An open question is whether the xKS ansatz may resolve the above-mentioned problems
and lead to these so far unknown exact solutions.

Another significant reason for studying xKS spacetimes is that some of the known
exact solutions can be cast to the xKS form. For instance, metrics with vanishing
scalar invariants (VSI) \cite{ColeyMilsonPravdaPravdova2004,ColeyFusterHervikPelavas2006}
are examples of xKS spacetimes as will be pointed out in section
\ref{sec:xKS:Kundt}. Some expanding xKS spacetimes representing black holes are
also known, namely the Chong, Cveti\v{c}, L\"u, and Pope (CCLP) solution
\cite{ChongCveticLuPope2005} discussed in section \ref{sec:xKS:CCLP}. In fact,
the investigation of the CCLP metric in \cite{AlievCiftci2008} has led to the
introduction of the notion of the xKS form.

Lastly, the xKS metric seems to be still sufficiently simple to be treated
analytically with occasional help of the computer algebra system {\sc Cadabra}
\cite{Peeters2006,Peeters2007} which allows to perform otherwise tedious symbolic
calculations in arbitrary unspecified dimension.

The paper is organized as follows. In section \ref{sec:xKS:general}, we analyze xKS
spacetimes with a general null Kerr--Schild vector field $\bk$, compare frames
in the full and background metrics and discuss the relation of the vectors $\bk$
and $\bm$ appearing in the xKS ansatz. In section \ref{sec:xKS:geodetic}, we assume
$\bk$ to be geodetic which allows us to express the Ricci and Riemann tensors
explicitly and, consequently, determine algebraic type of xKS spacetimes. Kundt
xKS metrics are examined in section \ref{sec:xKS:Kundt}. This assumption on
the geometry of the KS vector $\bk$ simplifies the subsequent calculations and
the possible Weyl type further specializes depending on the form of the scalar
function $\K$ and the Ricci tensor. We also provide examples of Kundt spacetimes
admitting the xKS form and discuss the relation of Ricci-flat \pp waves to the xKS
class. Finally, in section \ref{sec:xKS:CCLP}, the CCLP black hole as an example
of an expanding xKS spacetime is studied. For this solution, we establish a null
frame, determine the Weyl type and show that the optical matrix obeys the optical
constraint.

\subsection{Preliminaries}

Throughout the paper, we assume that the dimension $n \geq 4$ and that the background
metric $\bar{g}_{ab}$ representing an (anti-)de Sitter or Minkowski spacetime
takes the conformally flat form
\begin{equation}
  \bar{g}_{ab} = \Omega \eta_{ab}, \qquad
  \eta_{ab} \, \d x^a \d x^b = -\d t^2 + \d x_1^2 + \ldots + \d x_{n-1}^2
  \label{xKS:bgmetric}
\end{equation}
with the corresponding conformal factor
\begin{equation}
  \Omega_\mathrm{AdS} = \frac{(n-2)(n-1)}{2\Lambda t^2}, \qquad
  \Omega_\mathrm{dS} = - \frac{(n-2)(n-1)}{2\Lambda {x_1}^2},
  \label{xKS:Omega(A)dS}
\end{equation}
or $\Omega = 1$ in the case of flat background, respectively. The cosmological
constant $\Lambda$ is defined such that the vacuum Einstein field equations read
\begin{equation}
  R_{ab} = \frac{2 \Lambda}{n-2} g_{ab.}
  \label{xKS:Einstein-spaces}
\end{equation}

In $n$-dimensional spacetimes, it is convenient to introduce a real null frame
consisting of two null vectors $\bn \equiv \bm^{(0)}$, $\bl \equiv \bm^{(1)}$
and $n - 2$ orthonormal spacelike vectors $\bm^{(i)}$ obeying
\begin{equation}
  \fl
  n^a n_a = \ell^a \ell_a = n^a m^{(i)}_a = \ell^a m^{(i)}_a = 0, \qquad
  n^a \ell_a = 1, \qquad
  m^{(i) a} m^{(j)}_a = \delta_{ij}.
  \label{intro:frame:constraints}
\end{equation}
The indices $a, b, \ldots$ range from 0 to $n - 1$, the indices $i, j, \ldots$
running from 2 to $n - 1$ number the spacelike frame vectors. Under boosts, the frame
transforms as
\begin{equation}
  \ell' = \lambda \ell, \qquad
  n' = \lambda^{-1} n, \qquad
  m'^{(i)} = m^{(i)}
  \label{intro:frame:boost}
\end{equation}
and we say that some quantity $q$ has a boost weight $w$ if it obeys
\begin{equation}
  q' = \lambda^w q.
\end{equation}

We adopt the notation of the higher dimensional Newman--Penrose formalism
\cite{ColeyMilsonPravdaPravdova2004,PravdaPravdovaColeyMilson2004,OrtaggioPravdaPravdova2007}.
Namely, the Ricci rotation coefficients $L_{ab}$, $N_{ab}$ and $\M{i}{bc}$ are
defined as frame components of covariant derivatives of the frame vectors
\begin{equation}
  \fl
  \ell_{a;b} = L_{cd} \, m_a^{(c)} m_b^{(d)}, \qquad
  n_{a;b} = N_{cd} \, m_a^{(c)} m_b^{(d)}, \qquad
  m_{a;b}^{(i)} = \M i {cd} \, m_a^{(c)} m_b^{(d)}
  \label{intro:Riccicoeff}
\end{equation}
and the directional derivatives along the corresponding frame vectors are denoted
as
\begin{equation}
  \D \equiv \ell^a \nabla_a, \qquad
  \T \equiv n^a \nabla_a, \qquad
  \delta_i \equiv m_{(i)}^a \nabla_a.
  \label{intro:frame:derivatives}
\end{equation}
The optical matrix $L_{ij}$ has a special geometrical meaning. If $\bl$ is
geodetic, $L_{ij}$ is invariant under null rotations with $\bl$ fixed and can be
decomposed as
\begin{equation}
  L_{ij} = S_{ij} + A_{ij}, \quad
  S_{ij} \equiv L_{(ij)} = \sigma_{ij} + \theta \delta_{ij}, \quad
  A_{ij} \equiv L_{[ij]},
  \label{intro:Lij}
\end{equation}
where the trace $\theta$, symmetric traceless part $\sigma_{ij}$, and anti-symmetric
part $A_{ij}$ are related to the expansion, shear, and twist of the geodetic
congruence $\bl$, respectively. The shear and twist scalars are defined as
\begin{equation}
  \sigma^2 \equiv \sigma_{ij} \sigma_{ij}, \qquad
  \omega^2 \equiv A_{ij} A_{ij}.
  \label{intro:optical_scalars}
\end{equation}

According to the form of the xKS ansatz, it is convenient to identify the null
and spacelike vectors $\bk$, $\bm$ appearing in the xKS metric \eqref{xKS:metric}
with the vectors $\bl$, $\bm^{(2)}$ of the null frame
\eqref{intro:frame:constraints}, respectively. The corresponding Ricci rotation
coefficients will be denoted as $L_{ab}$ and $M_{ab} \equiv \M{2}{ab}$. It is
also useful to define indices $\itilde, \jtilde, \ldots = 3, \ldots, n - 1$
denoted by tilde such that the vector $\bm \equiv \bm^{(2)}$ is excluded in
the notation $\bm^{(\itilde)}$.

%% file: xKS_general_KS_vector.tex
\section{General Kerr--Schild vector field}
\label{sec:xKS:general}

As in the case of the GKS ansatz \cite{MalekPravda2010}, first important result
follows from the boost weight 2 component of the Ricci tensor
$R_{00} = R_{ab} k^a k^b$. This is the simplest component since $\bk$ and $\bm$
obey \eqref{xKS:metric:km} and thus many terms of the Ricci tensor of the xKS
metric vanish when twice contracted with the null vector $\bk$. Although the
calculations are much more involved than in the case of GKS spacetimes, we obtain
a quite simple expression
\begin{eqnarray}
  \fl
  R_{00} = 2 \H L_{i0} L_{i0}
  	- \half \K^2 L_{\itilde 0} L_{\itilde 0}
	+ 2 \K L_{i(i} L_{2)0}
	+ \K L_{\itilde 0} M_{\itilde 0}
	+ 2 \D\K L_{20}
	+ \K \D L_{20} \nonumber \\
  	- \half (n - 2) \left( \frac{\Omega_{,ab}}{\Omega}
  	- \frac{3}{2} \frac{\Omega_{,a} \Omega_{,b}}{\Omega^2} \right) k^a k^b.
  \label{xKS:R00:general}
\end{eqnarray}
The last term vanishes identically since we assume the background metric $\bar g_{ab}$
to be a maximally symmetric vacuum for which the conformal factor is given by
\eqref{xKS:Omega(A)dS} or $\Omega = 1$ in the case of (A)dS or Minkowski spacetime,
respectively. Obviously, the component $R_{00}$ vanishes completely if $L_{i0} = 0$
and therefore
\begin{proposition}
  \label{xKS:proposition:geodetic_k}
  The boost weight 2 component of the Ricci tensor $R_{00} = R_{ab} k^a k^b$ vanishes
  if the null vector field $\bk$ in the extended Kerr--Schild metric \eqref{xKS:metric}
  is geodetic.
\end{proposition}
In the context of general relativity, proposition 1 implies that for a geodetic
$\bk$ the boost weight 2 component of the energy--momentum tensor vanishes $T_{00} = 0$
which holds not only if the energy--momentum tensor is absent, i.e.\ for Einstein
spaces, but also for spacetimes containing matter fields aligned with the Kerr--Schild
vector $\bk$ such as aligned Maxwell field $F_{ab} k^b \propto k_a$ or aligned
null radiation $T_{ab} \propto k_a k_b$. Note that in case $\K = 0$, the xKS ansatz
\eqref{xKS:metric} reduces to the GKS form and the implication of proposition
\ref{xKS:proposition:geodetic_k} becomes an equivalence \cite{MalekPravda2010}. 
For $\K \neq 0$, the condition is only sufficient and thus $R_{00}$ vanishes also
for a non-geodetic $\bk$ with a special choice of $\H$, $\K$ and $\bm$.

\subsection{Kerr--Schild congruence in the background}

One may easily relate the null frames in the full and background spacetimes.
Starting with the full metric $g_{ab}$ expressed in terms of the null frame
\eqref{intro:frame:constraints} as
\begin{equation}
  g_{ab} = 2 k_{(a} n_{b)} + m_a m_b
    + \delta_{\tilde{\imath}\tilde{\jmath}} \, m_a^{(\tilde{\imath})} m_b^{(\tilde{\jmath})},
\end{equation}
we compare this form with the xKS ansatz \eqref{xKS:metric}. Obviously, if we
set 
\begin{equation}
  \bar{n}_a = n_a + \H k_a + \K m_a,
  \label{xKS:background_n}
\end{equation}
then $\bar{g}_{ab} = 2 k_{(a} \bar{n}_{b)} + \delta_{ij} \, m_a^{(i)} m_b^{(j)}$
and thus $\bar{\bn}$, $\bk$, $\bm$, $\bm^{(\tilde{\imath})}$ form a null frame
in the background spacetime. The indices of the vectors $\bk$ and $\bm^{(\itilde)}$
can be raised and lowered by both metrics $g_{ab}$ and $\bar{g}_{ab}$, however,
one has to treat the vectors $\bar{\bn}$ and $\bm$ carefully since
\begin{equation}
  \bar{n}^a \equiv g^{ab} \bar{n}_b = n^a + \H k^a + \K m^a, \qquad
  m^a \equiv g^{ab} m_b,
\end{equation}
while
\begin{equation}
  \bar{g}^{ab} \bar{n}_a = n^b - \H k^b, \qquad
  \bar{g}^{ab} m_a = m^b - \K k^b.
  \label{xKS:m:raised_in_bg}
\end{equation}
The covariant derivative compatible with the background metric $\bar{g}_{ab}$
can be simply expressed using the covariant derivative compatible with the full
metric $g_{ab}$ and setting $\H = \K = 0$. The relation between the covariant
derivatives allows us to compare the Ricci rotation coefficients constructed in
the full spacetime using the frame $\bn$, $\bk$, $\bm$, $\bm^{(\itilde)}$ with
those ones in the background spacetime denoted by barred letters and expressed
in terms of the frame $\bar{\bn}$, $\bk$, $\bm$, $\bm^{(\itilde)}$
\begin{eqnarray}
  \fl
  L_{i0} = \bar L_{i0}, \qquad
  L_{10} = \bar L_{10}, \qquad
  L_{ij} = \bar L_{ij} - \K \delta_{2[i} \bar L_{j]0}, \label{xKS:Riccicoeff:1} \\
  \fl
  L_{1i} = \bar L_{1i}
    - \H \bar L_{i0}
    - \half \K \bar \Xi_i
    + \half \K \left( \bar L_{10}
    - \frac{\D\K}{\K}
    + \frac{\D \Omega}{\Omega} \right) \delta_{2i}, \\
  \fl
  L_{i1} = \bar L_{i1}
    - \half \K \bar \Xi_i
    - \half \K \left( \bar L_{10}
    + \frac{\D\K}{\K}
    - \frac{\D \Omega}{\Omega} \right) \delta_{2i}, \\
  \fl
  L_{11} = \bar L_{11}
    - \H \bar L_{10}
    - \D\H
    - \K \bar \Theta
    + \left( \H - \half \K^2 \right) \frac{\D\Omega}{\Omega}, \\
  \fl
  M_{\itilde 0} = \bar M_{\itilde 0} + \half \K \bar L_{\itilde 0}, \qquad
  M_{\itilde j} = \bar M_{\itilde j}
    + \K \bar L_{(\itilde j)}
    + \K \left( \bar L_{[2 \itilde]} + \bar M_{\itilde 0} \right) \delta_{2j}
    - \half \K \frac{\D\Omega}{\Omega} \delta_{\itilde j}, \\
  \fl
  M_{\itilde 1} = \bar M_{\itilde 1}
    - \H \bar L_{\itilde 2}
    + \left( \H - \half \K^2 \right) \bar \Xi_{\itilde}
    - \half \K \left( \H \bar L_{\itilde 0}
    + 2 \bar L_{(1 \itilde)}
    - \bar M_{\itilde 2} \right)
    + \half \delta_{\itilde} \K, \\
  \fl
  \M{\itilde}{\jtilde 0} = \bM{\itilde}{\jtilde 0}, \qquad
  \M{\itilde}{\jtilde 1} = \bM{\itilde}{\jtilde 1}
    + 2 \H \bar L_{[\itilde \jtilde]}
    + \H \bM{\itilde}{\jtilde 0}
    + \K \bar M_{[\itilde \jtilde]}, \\
  \fl
  \M{\itilde}{\jtilde k} = \bM{\itilde}{\jtilde k}
    + \K \left( \bar L_{[ij]} + \bM{\itilde}{\jtilde 0} \right) \delta_{2k}, \qquad
  N_{i0} = \bar N_{i0}
    - \half \K \bar \Xi_i
    - \half \left( \K \bar L_{10}
    + \D\K \right) \delta_{2i}, \\
  \fl
  N_{i1} = \bar N_{i1}
    + \delta_i \H
    + \H \left(2 \bar L_{1i} - \bar L_{i1} + \bar N_{i0} - \H \bar L_{i0} \right)
    - \K \left( \bar N_{2i} + \bar M_{i1} + \H \bar \Xi_i \right)
    - \T \K \delta_{2i}
    \nonumber \\
    \fl
    \qquad - \left[ \K \left( \bar L_{11}
    - \H \bar L_{10}
    - \K \bar \Theta
    - \half \frac{\T\Omega}{\Omega}
    - \K \frac{\delta_2 \Omega}{\Omega} \right)
    + \left( \H + \half \K^2 \right) \frac{\D\Omega}{\Omega} \right] \delta_{2i}, \\
  \fl
  N_{ij} = \bar N_{ij}
    + \H \bar L_{ji}
    - \K \bar M_{[ij]}
    - \K \left( \bar L_{i1} - \bar N_{i0} \right) \delta_{2j}
    - \left( \H \delta_{ij} + \half K^2 \delta_{2i} \delta_{2j} \right) \frac{\D\Omega}{\Omega} \nonumber \\
    \fl
    \qquad - \left( \K \left(2 \bar L_{(1k)}
    - \H \bar L_{k0}
    - \K \bar \Xi_k \right)
    + \delta_k \K \right) \delta_{2[i} \delta_{j]k}
    - \half \K \frac{\delta_2 \Omega}{\Omega} \delta_{ij}
  \label{xKS:Riccicoeff:2}
\end{eqnarray}
with $\bar{\Xi}_i = \bar{L}_{2i} + \bar{M}_{i0}$ and $\bar \Theta = \bar L_{21} - \bar N_{20}$.
It follows from \eqref{xKS:Riccicoeff:1} that the null vector $\bk$ is
geodetic (and affinely parametrized) in the full spacetime $g_{ab}$ if and only
if it is geodetic (and affinely parametrized) in the background spacetime
$\bar{g}_{ab}$. Subsequently, a geodetic $\bk$ has the same optical properties,
i.e.\ the expansion, shear, and twist, in the both spacetimes since the optical
matrices $L_{ij}$ and $\bar{L}_{ij}$ are equal.

\subsection{Relation of the vector fields $\bk$ and $\bm$}
\label{sec:xKS:relation_k_m}

Inspired by the observation that the congruences $\bk$ and $\hat{\bm} = \zeta \bm$
of the CCLP black hole (discussed in section \ref{sec:xKS:CCLP}) obey
\eqref{xKS:CCLP:k_m}, let us study the consequences of the covariant condition
\begin{equation}
%  (m_{a;b} - m_{b;a}) k^b = - \frac{\D\zeta}{\zeta} m_a, \qquad
  (\zeta m_a)_{;b} k^b = (\zeta m_b)_{;a} k^b, \qquad
  k_{a;b} m^b = k_{b;a} m^b,
  \label{xKS:relation_k_m}
\end{equation}
restricting the geometry of the vectors $\bk$ and $\bm$. The contractions of
the first equation with the vectors $\bn$, $\bm$, $\bm^{(\itilde)}$ and the second
equation with $\bk$, $\bn$, $\bm^{(\itilde)}$ give
\begin{equation}
  L_{21} - N_{20} = 0, \qquad
  L_{22} = - \frac{\D\zeta}{\zeta}, \qquad
  L_{2 \itilde} + M_{\itilde 0} = 0
  \label{xKS:relation_k_m:Riccicoeff1}
\end{equation}
and
\begin{equation}
  L_{20} = 0, \qquad
  L_{[12]} = 0, \qquad
  L_{[2 \itilde]} = 0,
  \label{xKS:relation_k_m:Riccicoeff2}
\end{equation}
respectively. The Lie derivative of a one-form $\omega$ along a vector field $X$
under the condition $X^a \omega_a = 0$ reads
\begin{equation}
  (\mathcal{L}_{X} \omega)_a = \omega_{a;b} X^b + X^b_{\phantom{b};a} \omega_b
    = (\omega_{a;b} - \omega_{b;a}) X^b
\end{equation}
and therefore the relations \eqref{xKS:relation_k_m} can be equivalently
expressed as
\begin{equation}
  \mathcal{L}_{\bk} m_a = - \frac{\D\zeta}{\zeta} m_a, \qquad
  \mathcal{L}_{\boldsymbol{m}} k_a = 0.
  \label{xKS:relation_k_m:lie}
\end{equation}

It turns out that the additional covariant condition \eqref{xKS:relation_k_m}
restricts the geometry of the vectors $\bk$ and $\bm$ in the xKS metric \eqref{xKS:metric}
so that the implication of proposition \ref{xKS:proposition:geodetic_k} becomes
an equivalence. Thus, due to \eqref{xKS:relation_k_m:Riccicoeff1} and
\eqref{xKS:relation_k_m:Riccicoeff2}, all the terms apart from the first two
in \eqref{xKS:R00:general} vanish
\begin{equation}
  R_{00} = \left( 2\H - \half \K^2 \right) L_{\itilde 0} L_{\itilde 0},
  \label{xKS:R00:special}
\end{equation}
which immediately implies that 
\begin{corollary}
  \label{xKS:proposition:geodetic_k:2}
  In the special case when $\K^2 \neq 4 \H$,
  $\mathcal{L}_{\bm} k_a = 0$,
  and
  $\mathcal{L}_{\bk} m_a \propto m_a$,
  the null vector field $\bk$ in the extended
  Kerr--Schild metric \eqref{xKS:metric} is geodetic if and only if the boost
  weight 2 component of the Ricci tensor $R_{00} = R_{ab} k^a k^b$ vanishes.
\end{corollary}
On the other hand, the boost weight zero component $R_{00}$ vanishes regardless
of whether $\bk$ is geodetic in case $\K^2 = 4 \H$, $\mathcal{L}_{\bm} k_a = 0$,
and $\mathcal{L}_{\bk} m_a \propto m_a$.

Notice that the Lie bracket of the vectors $\bk$ and $\bm$ expressed in terms of
the Ricci rotation coefficients reads
\begin{equation}
  [\bm, \bk]^a = L_{20} \, n^a
    + (L_{12} + N_{20}) k^a
    + (L_{i2} - M_{i0}) m^a_{(i)}.
  \label{xKS:liebracket_k_m}
\end{equation}
If $\bk$ and $\bm$ satisfy the relation \eqref{xKS:relation_k_m}, then
\begin{equation}
  [\bm, \bk]^a =  2 L_{12} \, k^a
    + L_{22} \, m^a
    + 2 L_{2\itilde} \, m^a_{(\itilde)}
    \label{xKS:relation_k_m:liebracket}
\end{equation}
and therefore the vector fields $\bk$ and $\bm$ are surface-forming provided that
$L_{2 \itilde} = 0$. From \eqref{xKS:liebracket_k_m} it follows that general
$\bk$ and $\bm$ in Kundt xKS spacetimes are surface-forming for $M_{\itilde 0} = 0$
which is automatically satisfied if \eqref{xKS:relation_k_m} holds. Effectively,
$\zeta = 1$ in the relation \eqref{xKS:relation_k_m} for Kundt spacetimes since
\eqref{xKS:relation_k_m:Riccicoeff1} implies $\D\zeta = 0$. One may also consider
spacetimes with a recurrent null vector (RNV) field $\bl$ forming a subclass of Kundt
spacetimes characterized by the vanishing of all components $L_{ab}$ except for
$L_{11}$, see \ref{sec:appendix:Kundt}. In case that \eqref{xKS:relation_k_m} is
met, the vector fields $\bk \equiv \bl$ and $\bm$ of such RNV xKS spacetimes
commute, i.e.\ $[\bk, \bm] = 0$, and therefore the integral curves of $\bk$ and
$\bm$ can be chosen as coordinates.

To conclude this section, let us point out the compatibility of the relation
\eqref{xKS:relation_k_m} and the optical constraint \cite{OrtaggioPravdaPravdova2008,
OrtaggioPravdaPravdovaReall2012}
\begin{equation}
  L_{ik} L_{jk} = \frac{L_{lk} L_{lk}}{(n-2)\theta}S_{ij},
  \label{optical_constraint}
\end{equation}
which can be considered as a possible generalization of the
shear-free part of the Goldberg--Sachs theorem to higher dimensions, see
\cite{OrtaggioPravdaPravdova2012} for recent review. Although it has been shown
that any five-dimensional algebraically special Einstein spacetime obeys the
optical constraint \cite{OrtaggioPravdaPravdovaReall2012}, the situation in
dimension $n>5$ is not so clear. In fact, the optical constraint is satisfied
at least for certain classes of spacetimes in arbitrary dimension such as, for
instance, expanding Einstein GKS spacetimes \cite{MalekPravda2010}, type N
Einstein spacetimes and non-twisting type III Einstein spacetimes
\cite{PravdaPravdovaColeyMilson2004}.

The optical constraint \eqref{optical_constraint} implies that the optical matrix
$L_{ij}$ can be put to a block-diagonal form with $2\times2$ and $1\times1$ blocks
using an appropriate spins. For a geodetic Weyl aligned null direction (WAND)
$\bl$, the $r$-dependence of $L_{ij}$ satisfying the optical constraint
\eqref{optical_constraint} can be determined by integrating the Sachs equation
\cite{OrtaggioPravdaPravdova2007}. Thus, one gets
\cite{OrtaggioPravdaPravdova2008,MalekPravda2010}
\begin{equation}
  L_{ij} = \mathrm{diag} \left( \left[ \begin{array}{cc} s_1 & A_1 \\ - A_1 & s_1 \end{array} \right], \dots,
    \left[ \begin{array}{cc} s_p & A_p\\ - A_p & s_p \end{array} \right], \frac{1}{r}, \dots, \frac{1}{r}, 0, \dots, 0 \right),
  \label{optical_constraint:Lij}
\end{equation}
where
\begin{equation}
  s_\mu = \frac{r}{r^2 + a_\mu^2}, \qquad
  A_\mu = \frac{a_\mu}{r^2 + a_\mu^2}
\end{equation}
and $a_\mu$ are arbitrary functions not depending on $r$. Comparing
\eqref{xKS:relation_k_m:Riccicoeff2}, namely $L_{[2\itilde]} = 0$, with the optical
matrix \eqref{optical_constraint:Lij}, it follows that the vector $\bm$ must not
lie in any plane spanned by two spacelike frame vectors corresponding to a $2\times2$
block with non-vanishing twist of the null geodetic congruence $\bk$. Therefore,
omitting the degenerate case $L_{22} = 0$, $\bm$ lies in a $1\times1$ block of
the optical matrix, i.e.\ $L_{22} = r^{-1}$. From \eqref{xKS:relation_k_m:Riccicoeff1}
we then obtain $\zeta = \alpha r^{-1}$, where $\alpha$ does not depend on the
affine parameter $r$ along null geodesics $\bk$. It is shown in section
\ref{sec:xKS:CCLP} that for the CCLP black hole the vectors $\bk$ and $\bm$ satisfy
the relation \eqref{xKS:relation_k_m} and the optical constraint
\eqref{optical_constraint} also holds. In this case the function $\alpha$
corresponds to $\nu$.

%% file: xKS_geodetic_KS_vector.tex
\section{Geodetic Kerr--Schild vector field}
\label{sec:xKS:geodetic}

From now on, we assume that $\bk$ is the tangent vector field of a null geodetic
congruence and therefore $R_{00} = 0$ as follows from the proposition
\ref{xKS:proposition:geodetic_k}. Moreover, without loss of generality, the
geodesics are considered to be affinely parametrized. In terms of the Ricci
rotation coefficients, this means $L_{i0} = L_{10} = 0$ which considerably
simplifies the following calculations. Note that in the case $\K = 0$, i.e.\ for
GKS metrics, the assumption of Einstein spaces or spacetimes with aligned matter
fields $T_{ab} k^b \propto k_a$ in the context of general relativity implies that
$\bk$ is geodetic.

The frame components of the Ricci and Riemann tensors for xKS spacetimes
\eqref{xKS:metric} with a geodetic and affinely parametrized vector field $\bk$
are presented in \ref{sec:appendix:RicciRiemann}. All these components are much
more complicated than in the case of GKS spacetimes, in particular the boost
weight 1 components of the Ricci tensor $R_{0i}$ \eqref{xKS:R0i} and the Riemann
tensor $R_{010i}$ \eqref{xKS:R010i}, $R_{0ijk}$ \eqref{xKS:R0ijk}, respectively,
no longer vanish identically. However, since the boost weight 2 components
$R_{00}$ \eqref{xKS:R00} and $R_{0i0j}$ \eqref{xKS:R0i0j} are zero, it follows
that the same also holds for the Weyl tensor
\begin{equation}
  C_{0i0j} = 0
\end{equation}
and therefore
\begin{proposition}
  \label{xKS:proposition:Weyltype}
  Extended Kerr--Schild spacetimes \eqref{xKS:metric} with a geodetic
  Kerr--Schild vector $\bk$ are of the Weyl type I with $\bk$
  being the WAND.
\end{proposition}
In general, xKS spacetimes with a geodetic $\bk$ are not necessarily of Weyl type II
which confirms one of our motivations that these spacetimes may cover more general
algebraic types than Einstein GKS spacetimes.

Employing the components $R_{ij}$ of the Ricci tensor, one may show that non-expanding
Einstein GKS spacetimes belong to the Kundt class (i.e.\ $\theta = 0$ implies
$\sigma = \omega = 0$) and the optical matrix $L_{ij}$ of expanding Einstein GKS
spacetimes satisfies the optical constraint \cite{OrtaggioPravdaPravdova2008,
MalekPravda2010}. In fact, these results hold for more general class of GKS
spacetimes not necessarily Einstein \eqref{xKS:Einstein-spaces}, namely, it
suffices to assume $R_{ij} = \frac{2 \Lambda}{n-2} \delta_{ij}$ with $R_{01}$
being arbitrary.

In the case of xKS spacetimes, the optical properties of the null congruence $\bk$
are not so restricted, neither if one assumes the relation \eqref{xKS:relation_k_m}
between the vectors $\bk$ and $\bm$, and therefore non-expanding xKS spacetimes
may have, in principle, non-vanishing shear and twist. More precisely, if
a non-expanding geodetic null congruence $\bl$ of a spacetime with $R_{ab} \ell^a \ell^b = 0$
is non-shearing, it is consequently non-twisting and vice versa \cite{OrtaggioPravdaPravdova2007}.
For non-expanding xKS spacetimes, such a congruence is $\bk$ since $R_{00} = 0$
and thus xKS spacetimes with $\theta = 0$ are either Kundt or have both shear
and twist non-vanishing.

In case the vectors $\bk$ and $\bm$ satisfy the relation \eqref{xKS:relation_k_m}
restricting the geometry of xKS spacetimes, the frame components of the Ricci
and Riemann tensors further simplify. Then, due to \eqref{xKS:relation_k_m:Riccicoeff1}
and \eqref{xKS:relation_k_m:Riccicoeff2}, the boost weight 1 components of the
Ricci tensor \eqref{xKS:R0i}
\begin{eqnarray}
  R_{02} = - \half (L_{\itilde \itilde} + \D) (\D\K + \K L_{22})
    - \K \omega^2
    - \K L_{2 \itilde} L_{2 \itilde}
    + \K L_{22} L_{\itilde \itilde}, \\
    R_{0 \itilde} = (\D\K + \K L_{22}) L_{2 \itilde} - \K L_{2 \jtilde} L_{\jtilde \itilde} + \K L_{2 \itilde} L_{\jtilde \jtilde}
\end{eqnarray}
and the Riemann tensor \eqref{xKS:R010i}, \eqref{xKS:R0ijk}
\begin{eqnarray}
  R_{0102} = \frac{1}{2} \D (\D\K + \K L_{22}), \label{xKS:km:R0102} \\
  R_{010\itilde} = - \frac{1}{2} (\D\K + \K L_{22}) L_{2 \itilde}, \\
  R_{022\itilde} = - \frac{1}{2} (\D\K + \K L_{22}) L_{2 \itilde} + \K L_{2 \jtilde} A_{\jtilde \itilde}, \\
  R_{02\itilde\jtilde} = - (\D\K + \K L_{22}) A_{\itilde \jtilde} + \K L_{[\itilde |\tilde{k}} L_{\tilde{k} |\jtilde ]}, \\
  R_{0\itilde 2\jtilde} = - \frac{1}{2} (\D\K + \K L_{22}) L_{\itilde \jtilde} + \K L_{22} S_{\itilde \jtilde}
    - \K L_{2 \itilde} L_{2 \jtilde} + \K A_{\itilde \tilde{k}} L_{\tilde{k} \jtilde}, \\
  R_{0\itilde \jtilde \tilde{k}} = - 2\K S_{\itilde [\jtilde} L_{\tilde{k}] 2}, \label{xKS:km:R0IJK}
\end{eqnarray}
are given only in terms of the scalar function $\K$ and the optical matrix $L_{ij}$.
If all boost weight 1 components of the Riemann tensor
\eqref{xKS:km:R0102}--\eqref{xKS:km:R0IJK} vanish, all boost weight 1 components
of the Ricci and Weyl tensors vanish as well and, consequently, the spacetime is
of Weyl type II. Obviously, this holds for $\K = 0$, when the xKS metric
reduces to the GKS form studied in \cite{MalekPravda2010}. In the following, we
assume $\K \neq 0$ and split $L_{\itilde \jtilde}$ into its symmetric and anti-symmetric
part, respectively. The independent conditions for type II then read
\begin{eqnarray}
  \D \varkappa = 0,
  \label{xKS:km:typeII:Dkappa} \\
  \varkappa L_{\itilde} = 0,
  \label{xKS:km:typeII:Li} \\
  A_{\itilde \jtilde} L_{\jtilde} = 0,
  \label{xKS:km:typeII:AijLj} \\
  S_{\itilde [\jtilde} L_{\tilde{k}]} = 0,
  \label{xKS:km:typeII:SijLk} \\
  \varkappa A_{\itilde \jtilde} = 2 \K S_{[\itilde|\tilde{k}} A_{\tilde{k}|\jtilde]},
  \label{xKS:km:typeII:Aij} \\
  (2L_{22} - \varkappa \K^{-1}) S_{\itilde \jtilde} = 2 L_{\itilde} L_{\jtilde}
    - 2 A_{\itilde \tilde k} A_{\tilde k \jtilde} - 2 A_{(\itilde| \tilde k} S_{\tilde k |\jtilde)},
  \label{xKS:km:typeII:Sij}
\end{eqnarray}
where we denote $\varkappa \equiv \D\K + \K L_{22}$ and $L_{\itilde} \equiv L_{2\itilde}$
for convenience. In order to solve these equations, several cases have to be
investigated separately.

In the case $S_{\itilde \jtilde} = 0$, we consider two subcases. If $\varkappa = 0$,
contracting \eqref{xKS:km:typeII:Sij} with $L_{\jtilde}$ yields $L_{\itilde} L_{\jtilde} L_{\jtilde} = 0$
due to \eqref{xKS:km:typeII:AijLj} and thus $L_{\itilde} = 0$. Then, the trace of
\eqref{xKS:km:typeII:Sij} implies that the sum of squares of the elements of
$A_{\itilde \jtilde}$ vanish and therefore $A_{\itilde \jtilde} = 0$. Otherwise,
if $\varkappa \neq 0$, it follows directly from \eqref{xKS:km:typeII:Li},
\eqref{xKS:km:typeII:Aij} that $L_{\itilde} = 0$, $A_{\itilde \jtilde} = 0$ and
it remains to solve \eqref{xKS:km:typeII:Dkappa}.

In the case $S_{\itilde \jtilde}$ is of rank 1, $S_{\itilde \jtilde} = \mathrm{diag}(s_{(3)}, 0, \ldots, 0)$
and one gets from \eqref{xKS:km:typeII:SijLk} that $L_{\mu} = 0$, where
$\mu, \nu, \ldots = 4, \ldots, n - 1$. If $\varkappa = 0$, it follows from \eqref{xKS:km:typeII:Aij}
that $A_{3 \mu} = 0$. Taking $\itilde = \jtilde = \mu$ in \eqref{xKS:km:typeII:Sij}
and summing over $\mu$ leads to $A_{\mu \nu} = 0$. Therefore, $A_{\itilde \jtilde} = 0$
and $L_3$ is subject to $L_3^2 = L_{22} s_{(3)}$ as prescribed by \eqref{xKS:km:typeII:Sij}.
On the other hand, if $\varkappa \neq 0$, it immediately follows from \eqref{xKS:km:typeII:Li}
that $L_{\itilde} = 0$ and from \eqref{xKS:km:typeII:Aij} that $A_{\mu \nu} = 0$.
Putting $\itilde = 3$, $\jtilde = \mu$ to \eqref{xKS:km:typeII:Sij} yields
$A_{3 \mu} = 0$ and thus $A_{\itilde \jtilde} = 0$. Then \eqref{xKS:km:typeII:Sij}
implies that $L_{22} = \K^{-1} \D\K$, hence, $L_{22} \neq 0$ and it remains to satisfy
$\D^2 \K = 0$ with $\D \K \neq 0$.

In the case $m = \mathrm{rank}(S_{\itilde \jtilde}) \geq 2$, we can always set
$S_{\itilde \jtilde}$ to a diagonal form
$S_{\itilde \jtilde} = \mathrm{diag}(s_{(3)}, \ldots, s_{(m+2)}, 0, \ldots, 0)$
by appropriate rotations of the spacelike frame vectors $\bm^{(\itilde)}$ and
therefore \eqref{xKS:km:typeII:SijLk} leads to $L_{\itilde} = 0$. In the following,
it is convenient to employ indices $\alpha, \beta, \ldots = 3, \ldots, m+2$ and
$\mu, \nu, \ldots = m+3, \ldots, n-1$ such that $s_{(\alpha)} \neq 0$, $s_{(\mu)} = 0$.
Again, we consider two subcases with vanishing and non-vanishing $\varkappa$,
respectively.

If $m \geq 2$, $\varkappa = 0$, from \eqref{xKS:km:typeII:Aij} for $\itilde = \alpha$,
$\jtilde = \mu$, it follows that $A_{\alpha \mu} = 0$ and then, from
\eqref{xKS:km:typeII:Sij} for $\itilde = \mu$, $\jtilde = \nu$, one obtains
$A_{\mu \nu} = 0$. If, moreover, $L_{22} = 0$, \eqref{xKS:km:typeII:Sij} implies
for $\itilde = \jtilde = \alpha$ that $A_{\alpha \beta} = 0$. Therefore, $A_{\itilde \jtilde}$
vanishes, $S_{\itilde \jtilde}$ is an arbitrary diagonal matrix of rank $m$ and
$\D\K = 0$. If, otherwise, $L_{22} \neq 0$, necessarily $L_{22} s_{(\alpha)} \neq 0$ 
and \eqref{xKS:km:typeII:Sij} for $\itilde = \jtilde = \alpha$ yields
$L_{22} s_{(\alpha)} = \sum_{\beta} A^2_{\alpha \beta}$, which means that
$L_{22} s_{(\alpha)} > 0$ and for any given $\alpha$ there exists at least one
$\beta$ such that $A_{\alpha \beta} \neq 0$. However, \eqref{xKS:km:typeII:Aij}
with $\itilde = \alpha, \jtilde = \beta$ for non-vanishing $A_{\alpha \beta}$
implies $s_{(\alpha)} = - s_{(\beta)}$ and hence $L_{22} s_{(\beta)} < 0$, which
is a contradiction and the case $\varkappa = 0$, $L_{22} \neq 0$ is thus excluded.

If $m \geq 2$, $\varkappa \neq 0$, from \eqref{xKS:km:typeII:Aij} for $\itilde = \mu$,
$\jtilde = \nu$, we get $A_{\mu \nu} = 0$ and then \eqref{xKS:km:typeII:Sij} for
$\itilde = \mu$, $\jtilde = \nu$ gives $A_{\alpha \mu} = 0$. Now, if $L_{22} = \half \varkappa \K^{-1}$,
putting $\itilde = \jtilde = \alpha$ to \eqref{xKS:km:typeII:Sij} leads to
$A_{\alpha \beta} = 0$. Therefore, $A_{\itilde \jtilde}$ vanish, $s_{(\alpha)}$
are arbitrary and it remains to satisfy $\D^2\K = 0$. On the other hand, if $L_{22}
\neq \half \varkappa \K^{-1}$, the parts of \eqref{xKS:km:typeII:Aij} and
\eqref{xKS:km:typeII:Sij} involving $A_{\alpha \beta}$ can be written as
\begin{eqnarray}
  (\varkappa \K^{-1} - s_{(\alpha)} - s_{(\beta)}) A_{\alpha \beta} = 0,
  \label{xKS:km:typeII:1} \\
  (L_{22} - \half \varkappa \K^{-1}) s_{(\alpha)} = \sum_{\beta} A_{\alpha \beta} A_{\alpha \beta},
  \label{xKS:km:typeII:2} \\
  (s_{(\beta)} - s_{(\alpha)}) A_{\alpha \beta} = 2 \sum_{\gamma} A_{\alpha \gamma} A_{\beta \gamma},
  \qquad \alpha \neq \beta,
  \label{xKS:km:typeII:3}
\end{eqnarray}
where we do not use the summation convention over the repeated indices, instead
the summation symbol is explicitly indicated for clarity. Any row of the submatrix
$A_{\alpha \beta}$ contains at least one non-vanishing element since the left-hand
side of \eref{xKS:km:typeII:2} is non-vanishing. If we multiply \eqref{xKS:km:typeII:3}
by $A_{\alpha \beta}$ and sum over $\beta$, the right hand-side vanishes due to
the anti-symmetry of $A_{\beta \gamma}$ and we thus obtain
\begin{equation}
  \sum_{\beta} (s_{(\beta)} - s_{(\alpha)}) A_{\alpha \beta} A_{\alpha \beta} = 0.
  \label{xKS:km:typeII:4}
\end{equation}
Now, if a given row $\alpha$ contains just one non-vanishing element $A_{\alpha \beta}$,
then $s_{(\alpha)} = s_{(\beta)} = \half \varkappa \K^{-1}$ as follows from
\eqref{xKS:km:typeII:4} and \eqref{xKS:km:typeII:1}. If there are more
$A_{\alpha \beta_l} \neq 0$ with a fixed $\alpha$, then \eqref{xKS:km:typeII:1}
implies that all the corresponding $s_{(\beta_l)}$ are equal and
$s_{(\alpha)} = \varkappa \K^{-1} - s_{(\beta_l)}$. From \eqref{xKS:km:typeII:4},
one obtains that such $s_{(\beta_l)}$ are equal also to $s_{(\alpha)}$ and thus
$s_{(\alpha)} = \half \varkappa \K^{-1}$. Since this reasoning holds for any row of
$A_{\alpha \beta}$, all $s_{(\alpha)} = \half \varkappa \K^{-1}$. Therefore,
\eqref{xKS:km:typeII:3} reduces to $\sum_{\gamma} A_{\alpha \gamma} A_{\beta \gamma} = 0$
pointing out that the row vectors are orthogonal, which along with \eqref{xKS:km:typeII:2}
finally leads to $A_{\alpha \beta} = (\half \varkappa \K^{-1}(L_{22}
- \half \varkappa \K^{-1}))^{\frac{1}{2}} \, O^{\mathrm{A}}_{\alpha \beta}$,
where $O^{\mathrm{A}}_{\alpha \beta}$ is an arbitrary anti-symmetric orthogonal
matrix. Since the determinant of any orthogonal matrix is 1 or $-1$ and the
determinant of any regular $m \times m$ anti-symmetric matrix is positive for
even $m$ and vanishes for odd $m$, the dimension of the submatrix $A_{\alpha \beta}$
corresponding to the rank of $S_{\itilde \jtilde}$ has to be even. Note that
$S_{\alpha \beta}$ is a multiple of the identity matrix and thus commute with
$A_{\alpha \beta}$, as a consequence, one may simultaneously retain $S_{\alpha \beta}$
in the diagonal form $S_{\alpha \beta} = \half \varkappa \K^{-1} \delta_{\alpha \beta}$
and put $A_{\alpha \beta}$ to a block diagonal form consisting of $2 \times 2$ blocks
\begin{equation}
  A_{\alpha \beta} = \frac{\sqrt{\varkappa \K^{-1}(2 L_{22} - \varkappa \K^{-1})}}{2} \,
  \mathrm{diag} \! \left( \left[ \begin{array}{cc} 0 & 1 \\ -1 & 0 \end{array} \right], \ldots,
    \left[ \begin{array}{cc} 0 & 1 \\ -1 & 0 \end{array} \right] \right)
\end{equation}
by appropriate rotations of the spacelike frame vectors $\bm^{(\itilde)}$.

For all the given forms of the optical matrix solving
\eqref{xKS:km:typeII:Li}--\eqref{xKS:km:typeII:Sij}, one can integrate the Sachs
equation \cite{OrtaggioPravdaPravdova2007} following from the Ricci identities
which for xKS spacetimes with a geodetic $\bk$ reads
\begin{equation}
  \D L_{ij} = - L_{kj} \M{k}{i0} - L_{ik} \M{k}{j0} - L_{ik} L_{kj}.
\end{equation}
Therefore, we are able to determine the $r$-dependence of these optical matrices
and consequently of the corresponding functions $\K$ as follows
\begin{eqnarray}
  \fl
  L^{(1)}_{ij} = 0, \qquad \K^{(1)} = c_1 r + c_2,
  \label{xKS:km:typeII:form1} \\
  \fl
  L^{(2)}_{ij} = \mathrm{diag} \! \left(\frac{1}{r}, 0, \ldots, 0 \right), \qquad \K^{(2)} = c_1 r + \frac{c_2}{r},
  \label{xKS:km:typeII:form2} \\
  \fl
  L^{(3)}_{ij} = \frac{1}{1 + c^2_1 r^2} \, \mathrm{diag} \! \left(
    \left[ \begin{array}{cc} \frac{1}{r} & c_1 \\ c_1 & c^2_1 r \end{array} \right], 0, \ldots, 0 \right),
    \qquad \K^{(3)} = \frac{\sqrt{1 + c^2_1 r^2}}{c_2 r}, \qquad c_1 \neq 0,
  \label{xKS:km:typeII:form3} \\
  \fl
  L^{(4)}_{ij} = \mathrm{diag} \! \left(0, \frac{1}{r}, \frac{1}{r + c_2}, \ldots, \frac{1}{r + c_p}, 0, \ldots, 0 \right),
    \quad \mathrm{rank}(L^{(4)}_{ij}) \geq 1, \qquad \K^{(4)} = c_1,
  \label{xKS:km:typeII:form4} \\
  \fl
  L^{(5)}_{ij} = \mathrm{diag} \! \left(\frac{1}{r}, \frac{1}{r + c_2}, \ldots, \frac{1}{r + c_p}, 0, \ldots, 0 \right),
    \quad \mathrm{rank}(L^{(5)}_{ij}) \geq 2, \qquad \K^{(5)} = c_1 r,
  \label{xKS:km:typeII:form5} \\
  \fl
  L^{(6)}_{ij} = \mathrm{diag} \! \left( \frac{1}{r}, \mathcal{M}, \ldots, \mathcal{M} \right),
    \qquad \K^{(6)} = c_1 r + \frac{c_2}{r}, \qquad (c_1 \neq 0) \land (c_2 \neq 0),
  \label{xKS:km:typeII:form6}
\end{eqnarray}
respectively, where the arbitrary functions $c_\mu$ independent of $r$ are subject
to $\K \neq 0$ and
\begin{equation}
  \mathcal{M} = \left[ \begin{array}{cc} s & A \\ -A & s \end{array} \right], \qquad
  s = \frac{r}{r^2 + \frac{c_2}{c_1}}, \qquad A = \sqrt{\frac{c_2}{c_1}}\frac{1}{r^2 + \frac{c_2}{c_1}}.
\end{equation}
We thus arrive at
\begin{proposition}
  In the case that $\K \neq 0$, $\mathcal{L}_{\bm} k_a = 0$, and $\mathcal{L}_{\bk} m_a \propto m_a$, 
  extended Kerr--Schild spacetimes with a geodetic $\bk$ satisfying $R_{0i} = 0$
  are algebraically special if and only if the optical matrix $L_{ij}$ and the function $\K$
  can be put to any of the forms \eqref{xKS:km:typeII:form1}--\eqref{xKS:km:typeII:form6}.
\end{proposition}

Note that the case \eqref{xKS:km:typeII:form1} belongs to the Kundt class.
The optical matrix \eqref{xKS:km:typeII:form3} can be set to the form
$L_{ij} = \mathrm{diag}(\frac{1}{r}, 0, \ldots, 0)$ by an appropriate rotation
in the plane spanned by $\bm$ and $\bm^{(3)}$, therefore, the optical constraint
holds for the cases \eqref{xKS:km:typeII:form1}--\eqref{xKS:km:typeII:form3} and
\eqref{xKS:km:typeII:form6}. The optical matrices \eqref{xKS:km:typeII:form4} and
\eqref{xKS:km:typeII:form5} satisfy the optical constraint only for $c_\mu = 0$
where $\mu = 2,\ldots,p$ and if, moreover, the optical matrix \eqref{xKS:km:typeII:form5}
is regular such spacetimes belong to the Robinson--Trautmann class.
In four dimensions, the cases \eqref{xKS:km:typeII:form2}--\eqref{xKS:km:typeII:form3}
are excluded by the Goldberg--Sachs theorem.
In five dimensions, the cases \eqref{xKS:km:typeII:form2}--\eqref{xKS:km:typeII:form5}
with the corresponding optical matrices of rank 1 and 2, respectively, should be contained in the class of algebraically special non-twisting solutions
found in \cite{ReallGrahamTurner2012}.

%% file: xKS_kundt.tex
\section{Kundt extended Kerr--Schild spacetimes}
\label{sec:xKS:Kundt}

The simplest subclass of xKS spacetimes with a geodetic Kerr--Schild vector $\bk$
is characterized by the vanishing of the optical matrix $L_{ij}$. In other words,
the null geodetic congruence $\bk$ is non-expanding, non-shearing, and non-twisting,
i.e.\ such xKS spacetimes belong to the Kundt class. The relevant components for
the following analysis of the Ricci tensor \eqref{xKS:R0i}--\eqref{xKS:Rij} and
the Riemann tensor  \eqref{xKS:R010i}--\eqref{xKS:Rijkl} after substituting
$L_{ij} = 0$ reduce to
\begin{eqnarray}
  R_{0i} = - \half \D^2\K \delta_{2i}
    - \half \K \D M_{i0}
    - M_{i0} \D\K
    + \half \K M_{j0} \M{i}{j0},
  \label{xKS:Kundt:R0i} \\
  R_{01} = - \D^2\H
    + \half \K \D^2\K
    + \half (\D\K)^2
    - \half \delta_2 \D\K
    + \K^{-1} \D \! \left( \K^2 N_{20} \right) \nonumber \\
    \qquad - \half M_{ii} \D\K
    - \half \delta_i \! \left( \K M_{i0} \right)
    + \K M_{i0} N_{i0}
    - \half \K M_{i0} \M{i}{jj}
    + \frac{2\Lambda}{n-2},
  \label{xKS:Kundt:R01} \\
  R_{22} = - \delta_2 \D\K
    + \half (\D\K)^2
    + 2 L_{21} \D\K
    + \K M_{k2} M_{k0}
    + \frac{2 \Lambda}{n-2},
  \label{xKS:Kundt:R22} \\
  R_{\itilde 2} = - \half \delta_{\itilde} \D\K
    - \half M_{\itilde 2} \D\K
    - \half \delta_2 ( \K M_{\itilde 0} )
    + L_{\itilde 1} \D\K
    + \half \K M_{\itilde 0} \D\K \nonumber \\
    \qquad - \K \M{k}{(\itilde 2)} M_{k0}
    + 2 \K L_{21} M_{\itilde 0},
  \label{xKS:Kundt:RI2} \\
  R_{\itilde \jtilde} = - \D\K M_{(\itilde \jtilde)}
    - \delta_{(\itilde} ( \K M_{\jtilde) 0} )
    + 2 \K L_{(\itilde| 1} M_{|\jtilde) 0}
    - \K \M{k}{(\itilde \jtilde)} M_{k 0} \nonumber \\
    \qquad + \half \K^2 M_{\itilde 0} M_{\jtilde 0}
    + \frac{2 \Lambda}{n-2} \delta_{\itilde \jtilde}
  \label{xKS:Kundt:RIJ}
\end{eqnarray}
and
\begin{eqnarray}
  R_{010i} = \half \D^2\K \, \delta_{2i}
    + \half \K \D M_{i0}
    + M_{i0} \D\K
    - \half \K \M{i}{j0} M_{j0},
  \label{xKS:Kundt:R010i} \\
  R_{0ijk} = 0,
  \label{xKS:Kundt:R0ijk} \\
  R_{0101} = \D^2\H
    - \frac{1}{4} (\D\K)^2
    - \K M_{i0} N_{i0}
    + \D\! \left( \K L_{21} - \K N_{20} \right)
    - N_{20} \D\K \nonumber \\
    \qquad - \frac{1}{4} \K^2 M_{i0} M_{i0}
    - \frac{2 \Lambda}{(n - 2)(n - 1)},
  \label{xKS:Kundt:R0101} \\
  R_{01\itilde 2} = \half \delta_{\itilde}\D\K
    - \half \delta_2 \! \left( \K M_{\itilde 0} \right)
    - \half M_{\itilde 2} \D\K
    - \K M_{k0} \M{k}{[\itilde 2]},
  \label{xKS:Kundt:R01I2} \\
  R_{01\itilde\jtilde} = \delta_{[\itilde}\! \left( \K M_{\jtilde]0} \right)
    - M_{[\itilde\jtilde]} \D\K
    - \K M_{k0} \M{k}{[\itilde\jtilde]},
  \label{xKS:Kundt:R01IJ} \\
  R_{0212} = - \half \delta_2\D\K
    + \frac{1}{4} (\D\K)^2
    + L_{21} \D\K
    + \half \K M_{k0} M_{k2} \nonumber \\
    \qquad + \frac{2 \Lambda}{(n - 2)(n - 1)}, \\
  R_{021\itilde} = - \half \delta_{\itilde} \D\K
    + \frac{1}{4} \left( 2 L_{\itilde 1} + \K M_{\itilde 0} \right) \D\K
    + \half \K M_{\itilde 0} L_{21}
    + \half \K M_{k0} M_{k \itilde}, \\
  R_{0\itilde 12} = \frac{1}{4} \left( 2 L_{\itilde 1}
    - 2 M_{\itilde 2} + \K M_{\itilde 0} \right) \D\K
    - \half \delta_2 \! \left( \K M_{\itilde 0} \right) \nonumber \\
    \qquad + \half \K \left( M_{\itilde 0} L_{21}
    - M_{k0} \M{k}{\itilde 2} \right), \\
  R_{0\itilde 1 \jtilde} = - \half M_{\itilde\jtilde} \D\K
    - \half \delta_{\jtilde} \! \left( \K M_{\itilde 0} \right)
    + \K L_{(\itilde|1} M_{\jtilde)0}
    - \half \K M_{k0} \M{k}{\itilde \jtilde} \nonumber \\
    \qquad + \frac{1}{4} \K^2 M_{\itilde 0} M_{\jtilde 0}
    + \frac{2 \Lambda}{(n - 2)(n - 1)} \delta_{\itilde\jtilde}, \\
  R_{ijkl} = \frac{4 \Lambda}{(n-1)(n-2)} \delta_{i[k} \delta_{l]j},
  \label{xKS:Kundt:Rijkl}
\end{eqnarray}
respectively.

As follows from proposition \ref{xKS:proposition:Weyltype}, Kundt xKS metrics are
of Weyl type I, i.e.\ all boost weight 2 components of the Weyl tensor vanish.
The only non-trivial boost weight 1 components of the Riemann tensor
\eqref{xKS:Kundt:R010i} obey $R_{010i} = - R_{0i}$ and therefore the boost weight
1 components of the Weyl tensor are determined just by $R_{0i}$
\begin{equation}
  C_{0ijk} = \frac{1}{n - 2}(R_{0k} \delta_{ij} - R_{0j} \delta_{ik}), \qquad
  C_{010i} = \frac{3-n}{n-2} R_{0i}.
  \label{xKS:Kundt:Weyl:bw0}
\end{equation}
Obviously, Kundt xKS spacetimes are of Weyl type II if and only if $R_{0i} = 0$.
Note that the same statement holds even for general Kundt metrics not necessarily
of the xKS form \cite{OrtaggioPravdaPravdova2007}. Using \eqref{xKS:Kundt:R0i},
the equations $R_{02} = 0$ and $R_{0\itilde} = 0$ for xKS spacetimes take the forms
\begin{equation}
  \D^2\K = \K M_{\jtilde0} M_{\jtilde0}
  \label{xKS:Kundt:G02}
\end{equation}
and
\begin{equation}
  \D \! \left( \K^2 M_{\itilde 0} \right) = \K^2 M_{\jtilde0} \M{\itilde}{\jtilde0},
  \label{xKS:Kundt:G0I}
\end{equation}
respectively. The trivial solution $\K = 0$ corresponds to the GKS limit where
the components $R_{0i}$ identically vanish \cite{MalekPravda2010}. Since the Ricci
rotation coefficients $\M{i}{ja}$ are antisymmetric in the indices $i$ and $j$,
we eliminate the term on the right-hand side of \eqref{xKS:Kundt:G0I} by multiplying
the equation with $2 \K^2 M_{\itilde 0}$. The remaining term on the left-hand side
can be rewritten so that we arrive at $\D \! \left( \K^4 M_{\itilde 0} M_{\itilde 0} \right) = 0$
implying
\begin{equation}
  \K^4 M_{\itilde 0}M_{\itilde 0} = (c^0)^2,
  \label{xKS:Kundt:G0I:solution}
\end{equation}
where the function $c^0$ does not depend on the affine parameter $r$ along the null
geodesics of the Kerr--Schild congruence $\bk$. Substituting \eqref{xKS:Kundt:G0I:solution}
to \eqref{xKS:Kundt:G02}, we obtain $\K^3 \D^2\K = (c^0)^2$ determining the
$r$-dependence of the function $\K$ which has two distinct branches of solutions
\begin{eqnarray}
  \K = d^0 \sqrt{ (r + b^0)^2 + \frac{(c^0)^2}{(d^0)^4}}
    \qquad & \mbox{if $c^0 \neq 0$},
  \label{xKS:Kundt:K:rdep:c<>0} \\
  \K = f^0 r + e^0 & \mbox{if $c^0 = 0$},
  \label{xKS:Kundt:K:rdep:c=0}
\end{eqnarray}
where $b^0$, $d^0$, $e^0$ and $f^0$ are arbitrary functions not depending on $r$.
Since we assume $\K$ to be non-zero, $c^0$ vanishes if and only if all $M_{i0}$
vanish as can be seen directly from \eqref{xKS:Kundt:G0I:solution}. In the case
$c^0 \neq 0$ when $M_{\itilde 0} \neq 0$ one may determine the $r$-dependence
of $M_{\itilde 0}$. Without loss of generality, $\M{\itilde}{\jtilde 0}$ in
\eqref{xKS:Kundt:G0I} can be transform away using spatial rotations of $\bm^{(\itilde)}$
with $\bm^{(2)}$ fixed to obtain $\D \! \left( \K^2 M_{\itilde 0} \right) = 0$.
It is convenient to denote $\K^2 M_{\itilde 0} = (d^0)^2 \mu_{\itilde}$ and
comparing this form with \eqref{xKS:Kundt:K:rdep:c<>0} we can express $M_{\itilde 0}$
\begin{eqnarray}
  M_{\itilde 0} = \frac{\mu_{\itilde}}{(r + b^0)^2 + \mu_{\jtilde} \mu_{\jtilde}},
  \qquad \mu_{\itilde} \mu_{\itilde} = \frac{(c^0)^2}{(d^0)^4},
  \label{xKS:Kundt:Mi0:rdep}
\end{eqnarray}
where $\mu_{\itilde}$ does not depend on the affine parameter $r$. Finally, we
can conclude that
\begin{proposition}
  \label{xKS:proposition:Kundt:typeII}
  For Kundt extended Kerr--Schild spacetimes with $\K \neq 0$ and the vector
  field $\bk$ corresponding to the non-expanding, non-twisting, and non-shearing
  null geodesics the following statements are equivalent:
  \begin{itemize}
    \item[(i)] The boost weight 1 components $R_{0i} \equiv R_{ab} k^a m^b_{(i)} = 0$
      of the Ricci tensor vanish,
    \item[(ii)] the spacetime is of Weyl type II or more special,
    \item[(iii)] the function $\K$ takes the form \eqref{xKS:Kundt:K:rdep:c<>0} or
      \eqref{xKS:Kundt:K:rdep:c=0} along with $M_{i0}$ given by \eqref{xKS:Kundt:Mi0:rdep}.
  \end{itemize}
\end{proposition}
Note that in the context of the Einstein gravity, the vanishing of $R_{0i}$
corresponds to the vanishing of the components $T_{0i}$ of the energy--momentum
tensor.

We can proceed further and restrict type II Kundt xKS spacetimes to Weyl type III
by satisfying the additional conditions $C_{ijkl} = C_{01ij} = 0$. In terms of
the Riemann and Ricci tensors, these components of the Weyl tensor read
$C_{01ij} = R_{01ij}$ and
\begin{equation}
  C_{ijkl} = - \frac{2}{n-2} \left( \delta_{i[k} R_{l]j} + R_{i[k} \delta_{l]j} \right)
    + 2 \frac{R + 2 \Lambda}{(n-1)(n-2)} \delta_{i[k} \delta_{l]j},
  \label{xKS:Kundt:Cijkl}
\end{equation}
where we have substituted $R_{ijkl}$ for Kundt xKS spacetimes from
\eqref{xKS:Kundt:Rijkl}. Recall that $\Lambda$ represents a cosmological constant
of the background spacetime $\bar{g}_{ab}$.  Necessarily,
$R_{ij} = R_{(i)(i)} \delta_{ij} = \mathrm{diag}(R_{22}, R_{33}, \dots, R_{(n-1)(n-1)})$
for $C_{ijkl}$ to vanish and consequently
\begin{equation}
  R_{(i)(i)} + R_{(j)(j)} = \frac{R + 2 \Lambda}{n-1} \qquad \forall i,j: i \neq j.
  \label{xKS:Kundt:Cijkl:vanishing}
\end{equation}
For $n = 4$, it follows from \eqref{xKS:Kundt:Cijkl:vanishing} that $R_{22}$ and
$R_{33}$ have to satisfy
\begin{equation}
  \Lambda = - R_{01} + R_{22} + R_{33}.
  \label{xKS:Kundt:Cijkl:vanishing:n=4}
\end{equation}
In higher dimensions, \eqref{xKS:Kundt:Cijkl:vanishing} implies that
$R_{(i)(i)} = R_{(j)(j)}$ for all $i,j$ and
\begin{equation}
  \Lambda \delta_{ij} = \frac{n}{2} R_{ij} - R_{01} \delta_{ij}.
  \label{xKS:Kundt:Cijkl:vanishing:n>4}
\end{equation}

\begin{proposition}
  \label{xKS:proposition:Kundt:typeIII}
  Type II Kundt extended Kerr--Schild spacetimes with $\K \neq 0$ and $\bk$
  corresponding to the non-expanding, non-twisting, and non-shearing null geodesics
  are of Weyl type III if and only if the following two statements hold:
  \begin{itemize}
    \item[(i)] The boost weight zero components of the Ricci tensor
      $R_{ij}$ \eqref{xKS:Kundt:R22}--\eqref{xKS:Kundt:RIJ}
      are diagonal $R_{ij} = \mathrm{diag}(R_{22}, R_{33}, \dots, R_{(n-1)(n-1)})$
      and along with $R_{01}$ \eqref{xKS:Kundt:R01}
      satisfy either \eqref{xKS:Kundt:Cijkl:vanishing:n=4} in four dimensions or
      \eqref{xKS:Kundt:Cijkl:vanishing:n>4} for $n > 4$,
    \item[(ii)] the components of the Riemann tensor $R_{01ij}$ \eqref{xKS:Kundt:R01I2}
      and \eqref{xKS:Kundt:R01IJ} vanish.
  \end{itemize}
\end{proposition}
Note that the statement (i) is met obviously for Einstein spacetimes
$R_{ab} = \frac{2 \Lambda}{n-2} g_{ab}$ in arbitrary dimension.

Let us mention an example. We start with type II Kundt xKS spacetimes and set
$c^0 = f^0 = 0$, i.e.\ $M_{i0} = 0$ and $\K = e^0$. One can immediately see from
\eqref{xKS:Kundt:R01I2}, \eqref{xKS:Kundt:R01IJ} that $R_{01ij} = 0$ and from
\eqref{xKS:Kundt:R01}--\eqref{xKS:Kundt:RIJ} that
$R_{01} = - \D^2\H + \K \D N_{20} + \frac{2 \Lambda}{n-2}$ and
$R_{ij} = \frac{2 \Lambda}{n-2} \delta_{ij}$. If $\D^2\H = \K \D N_{20}$, both
statements of proposition \ref{xKS:proposition:Kundt:typeIII} are satisfied and
therefore such spacetimes are of Weyl type III.

The relation between the vectors $\bk$ and $\bm$ \eqref{xKS:relation_k_m} for
Kundt xKS spacetimes read
\begin{equation}
  L_{21} = N_{20}, \qquad
  D \zeta = 0, \qquad
  M_{\itilde 0} = 0, \qquad
  L_{[12]} = 0.
  \label{xKS:Kundt:relation_k_m}
\end{equation}
Since $M_{i0}$ vanishes, the relation is compatible only with $\K$ of the form
\eqref{xKS:Kundt:K:rdep:c=0}, i.e.\ a linear function of the affine parameter $r$,
in the case of type II subclass. Let us consider such spacetimes and assume
furthermore that $\D \K = 0$ which ensures type II. The situation is the same
as in the previous example, but now $\D N_{20} = 0$ due to the relation
\eqref{xKS:Kundt:relation_k_m} and the one of the Ricci identities  
\cite{OrtaggioPravdaPravdova2007} for Kundt spacetimes
$\D L_{21} = L_{i1} M_{i0} - C_{0102} + \frac{1}{n-2} R_{02}$. Therefore, these
spacetimes are of type III if $\D^2\H = 0$.

\subsection{Explicit examples of Kundt xKS metrics}

Here, we present explicit examples of Ricci-flat Kundt xKS spacetimes, namely
the class of spacetimes with vanishing scalar invariants (VSI)
\cite{ColeyMilsonPravdaPravdova2004,ColeyFusterHervikPelavas2006} which can be
described by metrics of the form
\begin{equation}
  \d s^2 = 2 \d u \, \d r 
    + 2 H(u, r, x^k) \, \d u^2
    + 2 W_i(u, r, x^k) \, \d u \, \d x^i
    + \delta_{ij} \, \d x^i \, \d x^j.
  \label{xKS:VSI:metric}
\end{equation}
It is easy to see that the VSI metrics \eqref{xKS:VSI:metric} belong to the class
of xKS spacetimes \eqref{xKS:metric} with the flat background metric
$\bar{g}_{ab} \, \d x^a \, \d x^b = 2 \d u \, \d r + \delta_{ij} \, \d x^i \, \d x^j$
and the following identification
\begin{equation}
  \fl
  \H = - H, \qquad
  \K = - \sqrt{W_i W_i}, \qquad
  k_a \, \d x^a = \d u, \qquad
  m_a \, \d x^a = \frac{W_i \, \d x^i}{\sqrt{W_j W_j}}.
  \label{xKS:VSI:xKS_form}
\end{equation}
The operator $\D = \partial_r$ then just corresponds to the derivative with respect
to the affine parameter $r$ of the non-expanding, non-shearing, and non-twisting
null geodesics generated by the vector field $\bk$. 

It is known \cite{ColeyMilsonPravdaPravdova2004} that VSI spacetimes are of Weyl
type III and admit only negative boost weight components of the Ricci tensor,
i.e.\ $R_{00} = R_{0i} = R_{01} = R_{ij} = 0$. All VSI metrics with the Ricci
tensor of types N and O have been given explicitly in \cite{ColeyFusterHervikPelavas2006}.
The VSI class can be divided into two distinct subclasses with vanishing and
non-vanishing quantity $L_{1i} L_{1i}$ denoted as $\epsilon = 0$ and $\epsilon = 1$,
respectively, which differ by the canonical choices of the functions $W_i$. The
subclass $\epsilon = 0$ corresponds to RNV spacetimes containing also \pp waves,
see \ref{sec:appendix:Kundt}. Assuming a particular algebraic type of
the Weyl and Ricci tensors, the functions $W_i$ and $H$ are further constrained
\cite{ColeyFusterHervikPelavas2006}.

Note that for VSI spacetimes the statements (i) and (ii) of proposition
\ref{xKS:proposition:Kundt:typeII} are clearly satisfied and therefore the function
$\K$ takes one of the forms \eqref{xKS:Kundt:K:rdep:c<>0} or
\eqref{xKS:Kundt:K:rdep:c=0} depending on $W_i(u, r, x^k)$. In the subclass
$\epsilon = 0$, the functions $W_i$ are given by
\cite{ColeyFusterHervikPelavas2006}
\begin{equation}
  W_2 = 0, \qquad
  W_{\itilde} = W_{\itilde}^0(u, x^k),
\end{equation}
where $W_{\itilde}^0$ satisfy some additional constraints and are independent
on the coordinate $r$ corresponding to an affine parameter along the geodetic
integral curves of the null vector $\bk$. It follows from \eqref{xKS:VSI:xKS_form}
that $\K$ is of the form \eqref{xKS:Kundt:K:rdep:c=0} with $f^0 = 0$ and does not
depend on $r$
\begin{equation}
  \K = - \sqrt{W_{\itilde}^0 W_{\itilde}^0} = e^0.
\end{equation}
Therefore $M_{i0}$ vanish and if also $N_{20} = 0$, then the vector $\bm$ is
parallelly transported along the null geodesics $\bk$. In fact, non-vanishing
$N_{20}$ can be always transformed away, while $M_{i0}$ remain unaffected, by
a null rotation with $\bk$ fixed \cite{OrtaggioPravdaPravdova2007} setting
$\D z_2 = - N_{20}$. Although this Lorentz transformation changes the vector
$\bm$ as $\hat\bm = \bm - z_2 \bk$, the xKS form \eqref{xKS:metric} of the metric
is preserved if we introduce a new function $\H$ such that $\hat\H = \H + z_2 \K$.

The subclass $\epsilon = 1$ of VSI spacetimes is characterized by the canonical
form of the functions $W_i(u, r, x^k)$ 
\cite{ColeyFusterHervikPelavas2006}
\begin{equation}
  W_2 = - \frac{2}{x^2} r, \qquad
  W_{\itilde} = W_{\itilde}^0.
  \label{xKS:VSI:W:e=1}
\end{equation}
In the special case where all $W^0_{\itilde}$ vanish, the function $\K$ corresponds
to \eqref{xKS:Kundt:K:rdep:c=0} with $e^0 = 0$
\begin{equation}
  \K = - \frac{2}{|x^2|} r = f^0 r.
\end{equation}
As in the previous case, $M_{\itilde0}$ = 0 and $N_{20}$ vanishes or can be set
to zero, consequently, the vector $\bm$ is parallelly transported along $\bk$.
On the other hand, if at least one of $W_{\itilde}$ in \eqref{xKS:VSI:W:e=1} is
non-zero, the function $\K$ takes the form \eqref{xKS:Kundt:K:rdep:c<>0}
\begin{equation}
  \K = - \sqrt{\frac{4}{(x^2)^2} r^2 + W_{\itilde}^0 W_{\itilde}^0}.
  \label{xKS:VSI:W:e=1:K}
\end{equation}
Comparing \eqref{xKS:Kundt:K:rdep:c<>0} with \eqref{xKS:VSI:W:e=1:K}, it immediately
follows that $b^0 = 0$, $d^0 = - \frac{2}{|x^2|}$, and
$(c^0)^2 = \frac{(x^2)^2}{4} W^0_{\itilde} W^0_{\itilde}$.
Now, the vector $\bm$ is not parallelly transported along $\bk$ since
$M_{\itilde 0} M_{\itilde 0} = \K^{-4} (c^0)^2 \neq 0$.

\subsection{Relation of higher dimensional \pp waves and the class of xKS spacetimes}

In the previous section, we have shown that all VSI metrics admit the xKS form
\eqref{xKS:metric}. The question is whether also all \pp waves belong to the class
of xKS spacetimes. In our discussion, we restrict ourselves to Einstein spaces,
i.e.\ vacuum solutions with a possible cosmological constant in the framework
of general relativity.

Higher dimensional \pp waves are, in general, of Weyl type II and Einstein \pp waves
are necessarily Ricci-flat as discussed in \ref{sec:appendix:Kundt}. Furthermore,
it is known that Kundt spacetimes of Weyl type III with the Ricci tensor of type
III, including Weyl type III Ricci-flat \pp waves, belong to the VSI class
\cite{ColeyMilsonPravdaPravdova2004}. Therefore, it remains to investigate
Ricci-flat \pp waves of genuine type II.

Any \pp wave spacetime can be described by a metric \cite{Brinkmann1925}
\begin{equation}
  \d s^2 = 2 \d u \left[ \d v + H(u, x^k) \, \d u + W_{i}(u, x^k) \, \d x^i \right]
    + g_{ij}(u,x^k) \, \d x^i \, \d x^j,
  \label{xKS:Kundt:ppwave}
\end{equation}
which cannot be cast to the xKS form for any arbitrary transverse metric $g_{ij}$.
However, the transverse Riemannian metric $g_{ij}$ of vacuum \pp waves is
Ricci-flat \cite{PodolskyZofka2008} and in the case of Kundt CSI metrics the
transverse Riemannian space is locally homogeneous \cite{ColeyHervikPelavas2005}.  
Since a Ricci-flat locally homogeneous Riemannian space is flat
\cite{PruferTricerriVanhecke1996}, we can conclude that Ricci-flat CSI \pp wave
metrics can be written in the form \eqref{xKS:Kundt:ppwave} with flat transverse
space, i.e.\ $g_{ij} = \delta_{ij}$, and thus belong to the class of xKS spacetimes
with Minkowski background.

Recall also that type N Ricci-flat Kundt spacetimes and consequently vacuum type N
\pp waves can be cast to the KS form \eqref{KS:KSansatz} as was shown in
\cite{OrtaggioPravdaPravdova2008}. All the above mentioned properties of higher
dimensional vacuum \pp waves are summarized in table \ref{tab:xKS:ppwaves}.
On the other hand, the situation in four dimensions is much more simple since
all vacuum \pp wave metrics are only of Weyl type N, belong to
the VSI class, and take the KS form.

\begin{table}
  \caption{\label{tab:xKS:ppwaves}Properties of higher dimensional Ricci-flat
  \pp waves. With regard to the particular algebraic type, such \pp waves belong
  to the various classes of spacetimes.}
  \begin{indented}
    \item[]\begin{tabular}{cccc}
      \br
      Weyl type & KS & xKS & VSI \\
      \mr
      N      & \checkmark & \checkmark & \checkmark \\
      III=III(a) & $\times$   & \checkmark & \checkmark \\
      II=II(abd) & $\times$   & CSI   & $\times$   \\
      \br
    \end{tabular}
  \end{indented}
\end{table}

%% file: xKS_expanding.tex
\section{Examples of expanding extended Kerr--Schild spacetimes}
\label{sec:xKS:CCLP}

In this section, we give an explicit example of expanding xKS metric, namely the
CCLP solution \cite{ChongCveticLuPope2005}. For such a spacetime, we construct
a null frame, show that the optical matrix obeys the optical constraint, and
determine algebraic type of the Weyl tensor. The CCLP metric represents a charged
rotating black hole in five-dimensional minimal gauged supergravity or equivalently
in the Einstein--Maxwell--Chern--Simons theory with a negative cosmological
constant $\Lambda$ and the Chern--Simons coefficient $\chi = 1$ described by the
field equations
\begin{eqnarray}
  R_{ab} = \frac{2}{3} \Lambda g_{ab}
    + 2 (F_{ac} F_{b}^{\phantom{b}c} - \frac{1}{6} g_{ab} F_{cd} F^{cd}),
  \label{xKS:CCLP:FEs1} \\
  \nabla_b F^{ab} + \frac{\chi}{2\sqrt{3}\sqrt{-g}} \epsilon^{abcde} F_{bc} F_{de} = 0.
  \label{xKS:CCLP:FEs2}
\end{eqnarray}
Note that the CCLP metric solves \eqref{xKS:CCLP:FEs1} and \eqref{xKS:CCLP:FEs2}
also for a positive cosmological constant.

In fact, the xKS ansatz has been first proposed in \cite{AlievCiftci2008} by
showing that the CCLP black hole can be cast to the form
\begin{equation}
  g_{ab} = \bar g_{ab} - 2 \H k_a k_b - 2 \hat \K k_{(a} \hat m_{b)},
  \label{xKS:CCLP:original}
\end{equation}
where we distinguish $\hat{\K}$, $\hat{\bm}$ from $\K$, $\bm$ in our definition
of the xKS ansatz \eqref{xKS:metric} and \eqref{xKS:metric:km} since the vector
$\hat{\bm}$ is not normalized to unity. In terms of spheroidal coordinates, the
(Anti-)de Sitter background metric $\bar{g}_{ab}$ and the vectors $\bk$, $\hat{\bm}$ are given
by \cite{AlievCiftci2008}
\begin{eqnarray}
  \fl
  \bar{g}_{ab} \, \d x^a \, \d x^b = - \left( 1 - \lambda r^2 \right) \frac{\Delta}{\Xi_a \Xi_b} \, \d t^2
    - 2 \d r \left( \frac{\Delta}{\Xi_a \Xi_b} \, \d t - \frac{a \sin^2 \theta}{\Xi_a} \, \d \phi
    - \frac{b \cos^2 \theta}{\Xi_b} \, \d \psi \right) \nonumber \\
    + \frac{\rho^2}{\Delta} \, \d \theta^2
    + \frac{\left( r^2 + a^2 \right) \sin^2 \theta}{\Xi_a} \, \d \phi^2
    + \frac{\left( r^2 + b^2 \right) \cos^2 \theta}{\Xi_b} \, \d \psi^2, \\
  \fl
  k_a \, \d x^a = - \frac{\Delta}{\Xi_a \Xi_b} \, \d t
    + \frac{a \sin^2 \theta}{\Xi_a} \, \d \phi
    + \frac{b \cos^2 \theta}{\Xi_b} \, \d \psi,
  \label{xKS:CCLP:k} \\
  \fl
  \hat m_a \, \d x^a = \lambda a b \frac{\Delta}{\Xi_a \Xi_b} \, \d t
    + \frac{b \sin^2 \theta}{\Xi_a} \, \d \phi + \frac{a \cos^2 \theta}{\Xi_b} \, \d \psi,
  \label{xKS:CCLP:mhat}
\end{eqnarray}
where $\lambda = \frac{\Lambda}{6}$, $a$ and $b$ are spins, $Q$ corresponds to
charge, $r$ is the spheroidal radial coordinate, $\phi$, $\psi$, $\theta$ are
the angular coordinates with usual ranges $\phi \in \langle 0, 2 \pi)$,
$\psi \in \langle 0, 2 \pi)$, $\theta \in \langle 0, \pi \rangle$, respectively,
and
\begin{eqnarray}
  \rho^2 = r^2 + a^2 \cos^2 \theta + b^2 \sin^2 \theta,
  \qquad \Xi_a = 1 + \lambda a^2, \\
  \Xi_b = 1 + \lambda b^2,
  \qquad \Delta = 1 + \lambda a^2 \cos^2 \theta + \lambda b^2 \sin^2 \theta.
\end{eqnarray}
The functions $\H$, $\hat{\K}$ and the one-form gauge potential proportional to
the null Kerr--Schild vector $\bk$ then read
\begin{equation}
  \H = - \frac{M}{\rho^2} + \frac{Q^2}{2 \rho^4}, \qquad
  \hat \K = - \frac{Q}{\rho^2}, \qquad
  A = - \frac{\sqrt{3} Q}{2 \rho^2} \bk.
\end{equation}

In order to put the CCLP metric into the xKS form with a unit vector $\bm$, we
rescale the vector $\hat{\bm}$ and include its norm to the function $\K$ so that
\begin{equation}
  m_a \, \d x^a = \frac{\lambda a b r}{\nu} \frac{\Delta}{\Xi_a \Xi_b} \, \d t
    + \frac{b r \sin^2 \theta}{\nu \Xi_a} \d \phi
    + \frac{a r \cos^2 \theta}{\nu \Xi_b} \d \psi,
  \label{xKS:CCLP:m}
\end{equation}
\begin{equation}
  \K = - \frac{Q \nu}{r \rho^2},
  \label{xKS:CCLP:K}
\end{equation}
where we define, for convenience, $\nu^2 \equiv \rho^2 - r^2 = a^2 \cos^2\theta + b^2 \sin^2\theta$.

We can simply choose $\bk$ and $\bm$ as the null and spacelike frame vectors $\bl$
and $\bm^{(2)}$, respectively. The remaining frame vectors $\bn$, $\bm^{(3)}$
and $\bm^{(4)}$ have to be determined by solving the constraints
\eqref{intro:frame:constraints}. One may find easier to construct the frame in
the background spacetime since the metric $\bar{g}_{ab}$ is not so complex as
the full metric $g_{ab}$ and subsequently employ \eqref{xKS:background_n}.
Such a frame can be expressed, for instance, as
\begin{eqnarray}
  \fl
  k^a \, \partial_a = \partial_r, 
  \label{xKS:CCLP:frame:k} \\
  \fl
  m_{(2)}^a \, \partial_a = \frac{ab}{r \nu} \, \partial_t
    - \frac{Q \nu^2 + a b \rho^2}{r \rho^2 \nu} \, \partial_r
    + \frac{b}{r \nu} \, \partial_\phi
    + \frac{a}{r \nu} \, \partial_\psi, \\
  \fl
  m_{(3)}^a \, \partial_a = \frac{\sqrt{\Delta} \sin\theta \cos\theta}{\rho^2} \Bigg( \frac{a^2 - b^2}{\Delta} \, \partial_t
    + \frac{\Xi_a (r^2 + b^2) + \Xi_b (r^2 + a^2)}{\Pi} \, \partial_r \Bigg) \nonumber \\
    + \frac{\sqrt{\Delta}}{\rho^2} \Bigg( \frac{a \cos\theta}{\sin\theta} \, \partial_\phi
    - \frac{b \sin\theta}{\cos\theta} \, \partial_\psi
    + r \, \partial_\theta \Bigg), \\
  \fl
  m_{(4)}^a \, \partial_a = \frac{r \sqrt{\Delta}}{\nu \rho^2} \Bigg( \frac{b^2 - a^2}{\Delta} \sin\theta \cos\theta (\partial_t - \Delta \, \partial_r)
    - \frac{a \cos\theta}{\sin\theta} \, \partial_\phi
    + \frac{b \sin\theta}{\cos\theta} \, \partial_\psi
    + \frac{\nu^2}{r} \, \partial_\theta \Bigg), \\
  \fl
  n^a \, \partial_a = - \frac{1}{\rho^2} \Bigg( r^2 + \frac{a^2 \Xi_a \cos^2\theta - b^2 \Xi_b \sin^2\theta}{\Pi} \Bigg) \partial_t
    + \frac{1}{2 \rho^2} \Bigg( 2 M - \frac{Q^2}{\rho^2} + \lambda r^2 (r^2 + a^2 + b^2) \nonumber \\
%%%%%%%%
    - \frac{1}{\Pi^2} \Bigg(
    \rho^2 - 4 (2 r^2 + a^2 + b^2) \sin^2\theta \cos^2\theta
    + \lambda \nu^2 (\nu^2 - 4 (1 + \lambda) r^2) \nonumber \\
    - \lambda a^2 b^2 (\Delta (1 + \Delta) - \lambda r^2 (1 - 2 \cos^2\theta)^2) \nonumber \\ 
    - \lambda (a^2 + b^2)((4(a^2 + b^2 + \lambda a^2 b^2) + \lambda^2 r^2 \nu^2) \sin^2\theta \cos^2\theta - \nu^2 \Delta) \nonumber \\
    - \lambda r^2 (a^2 \cos^2\theta - b^2 \sin^2\theta)(\Xi_a + \Xi_b)(1 - 2 \cos^2\theta) \nonumber \\
    + 4 \lambda r^2 (a^2 \Xi_a \cos^6\theta + b^2 \Xi_b \sin^6\theta) \Bigg)
    \Bigg) \partial_r
    - \frac{a}{\rho^2} \Bigg( 1 - \lambda r^2 + \frac{\Delta \Xi_b}{\Pi} \Bigg) \partial_\phi \nonumber \\
    - \frac{b}{\rho^2} \Bigg( 1 - \lambda r^2 - \frac{\Delta \Xi_a}{\Pi} \Bigg) \partial_\psi
    - \frac{r \Delta (\Xi_a + \Xi_b) \sin\theta \cos\theta}{\rho^2 \Pi} \, \partial_\theta,
  \label{xKS:CCLP:frame:n}
\end{eqnarray}
where $\Pi = \Xi_a \cos^2\theta - \Xi_b \sin^2\theta$.

Having established the frame, we can straightforwardly calculate the Ricci rotation
coefficients \eqref{intro:Riccicoeff}. It turns out that $L_{a0} = 0$, i.e.\ the
Kerr--Schild congruence $\bk$ is geodetic and affinely parametrized, but only
$\bm^{(3)}$ and $\bm^{(4)}$ are parallelly transported along $\bk$ since
\begin{eqnarray}
  N_{20} = - Q \frac{\nu}{\rho^4}
  \label{xKS:CCLP:frame:N20}
\end{eqnarray}
is non-vanishing. Interestingly, the optical matrix
\begin{equation}
  L_{ij} = \left( \begin{array}{ccc} \frac{1}{r} & 0 & 0 \\
    0 & \frac{r}{\rho^2} & \frac{\nu}{\rho^2} \\
    0 & -\frac{\nu}{\rho^2} & \frac{r}{\rho^2}
  \end{array} \right)
  \label{xKS:CCLP:Lij}
\end{equation}
takes the same block-diagonal form as in the case of uncharged five-dimensional
Kerr-(A)dS black hole \cite{MalekPravda2010} and therefore $L_{ij}$ also satisfies
the optical constraint \eqref{optical_constraint}. 

One may also show that the vectors $\bk$ and $\hat{\bm}$ of the CCLP metric
satisfy
\begin{equation}
  (\hat m_{a;b} - \hat m_{b;a}) k^b = 0, \qquad
  (k_{a;b} - k_{b;a}) \hat m^b = 0,
  \label{xKS:CCLP:k_m}
\end{equation}
which holds not only in the full spacetime, but also in the background spacetime,
i.e.\ regardless whether we use the covariant derivative compatible with the full
metric $g_{ab}$ or with the background metric $\bar{g}_{ab}$. The relation
\eqref{xKS:CCLP:k_m} immediately implies that $\bk$ and $\bm$ met
\eqref{xKS:relation_k_m} with $\zeta = |\hat{\bm}| = \frac{\nu}{r}$. It can be
seen directly from \eqref{xKS:CCLP:Lij} that $L_{22} = \frac{1}{r}$, $L_{2 \itilde} = 0$
and since from \eqref{xKS:relation_k_m:Riccicoeff1}, \eqref{xKS:relation_k_m:Riccicoeff2}
and \eqref{xKS:CCLP:frame:N20} it follows that $L_{12} = - Q \frac{\nu}{\rho^4}$,
the Lie bracket \eqref{xKS:relation_k_m:liebracket} then read
\begin{equation}
  [\bk, \bm]^a = -2 Q \frac{\nu}{\rho^4} k^a + \frac{1}{r} m^a
\end{equation}
and therefore the vector fields $\bk$ and $\bm$ are surface-forming.

In agreement with proposition \ref{xKS:proposition:Weyltype}, the CCLP spacetime
is of Weyl type I since the boost weight 2 components of the Weyl tensor expressed
using the frame \eqref{xKS:CCLP:frame:k}--\eqref{xKS:CCLP:frame:n} completely
vanish
\begin{equation}
  C_{0i0j} = 0,
\end{equation}
however, we show that it is not more special. First, let us assume that the CCLP
spacetime is of type II, therefore, the Weyl tensor satisfies the corresponding
Bel--Debever criterion \cite{Ortaggio2009} for this type
\begin{equation}
  \ell_{[e} C_{a]b[cd} \ell_{f]} \ell^b = 0
  \label{xKS:CCLP:BelDeveber}
\end{equation}
and we look for the multiple WAND $\bl = \ell^t \partial_t + \ell^r \partial_r
+ \ell^\theta \partial_\theta + \ell^\phi \partial_\phi + \ell^\psi \partial_\psi$.
For simplicity, but without loss of generality, we consider only one spin $a$
to be non-zero. The component $\ell_{[r} C_{\theta]c[\psi r} \ell_{\theta]} \ell^c$
vanishes if $\ell^t = \frac{a \sin^2\theta}{\Delta} \ell^\phi$. Moreover,
$\ell_{[t} C_{\theta]c[\psi r} \ell_{\theta]} \ell^c = 0$ implies that
$\ell^\phi = - \frac{a \cos\theta}{r \sin\theta} \ell^\theta$ and also that either
$\ell^\theta \neq 0 \land \ell^\psi \neq 0$ and then $\ell^r$ can be expressed
in terms of $\ell^\theta$ and $\ell^\psi$, or $\ell^\theta = \ell^\psi = 0$ and
therefore $\ell = \ell^r \partial_r$. In the former case, $\bl$ cannot be a null
vector since $\ell_a \ell^a = \rho^4 \, {\ell^\theta}^2 + \Delta r^4 \cos^2\theta \, {\ell^\psi}^2$
is a sum of squares with positive coefficients. In the latter case,
$\ell_{[t} C_{\theta]c[\psi t} \ell_{\theta]} \ell^c = 0$ implies that $\ell^r = 0$
and consequently $\bl = 0$. Thus, we can conclude that there is no multiple WAND
$\bl$ satisfying the criterion \eqref{xKS:CCLP:BelDeveber} and the CCLP spacetime
is of genuine type I.

Only if either $\nu = 0$, i.e.\ non-rotating limit, or in
the uncharged case when the metric corresponds to the five-dimensional Kerr-(A)dS
black hole, the metric reduces to the GKS form since $\K$ \eqref{xKS:CCLP:K}
vanishes and the Weyl tensor is of type D.

Let us mention the results of \cite{PravdaPravdovaOrtaggio2007} that stationary
spacetimes with the metric remaining unchanged under reflection symmetry and with
non-vanishing expansion are of Weyl types G, I$_i$, D or conformally flat. The
CCLP metric obeys these conditions along with the reflection symmetry $t \rightarrow -t$,
$\phi \rightarrow -\phi$, $\psi \rightarrow -\psi$ of the metric in
Boyer--Lindquist-type coordinates given in \cite{ChongCveticLuPope2005}.
Therefore, more specifically, the CCLP solution belongs to the subtype I$_i$ of
Weyl type I. 

Finally, note also that we can adapt the frame \eqref{xKS:CCLP:frame:k}--\eqref{xKS:CCLP:frame:n}
to be parallelly transported along the null geodesics $\bk$ using an appropriate
null rotation with $\bk$ fixed. However, this operation changes $\bm^{(2)}$ and
thus breaks the identification $\bm \equiv \bm^{(2)}$. Such a Lorentz transformation
which set $N_{20}$ to zero is
\begin{equation}
  \fl
  \bk' = \bk, \qquad
  \bm' = \bm - z_2 \bk, \qquad
  \bm'^{(\itilde)} = \bm^{(\itilde)}, \qquad
  \bn' = \bn + z_2 \bm - \frac{1}{2} z_2^2 \bk,
  \label{xKS:CCLP:transformation}
\end{equation}
where
\begin{equation}
  z_2 = \frac{Q}{2 \nu^2} \left( \arctan \frac{r}{\nu} + \frac{\nu r}{\rho^2} \right).
\end{equation}
Obviously, the transformation \eqref{xKS:CCLP:transformation} preserves the xKS
form of the CCLP metric with $\bk'$ and $\bm'$ being the corresponding null and
spacelike unit vectors, respectively. The function $\K$ remains unchanged and
$\H$ is now taken as
\begin{equation}
  \H' = \H + z_2 \K = - \frac{1}{\rho^2} \left(M + \frac{Q^2}{2 \nu r} \arctan\frac{r}{\nu} \right).
\end{equation}

%% file: xKS_conclusion.tex
\section{Conclusion}

We have studied xKS spacetimes in any dimension $n \geq 4$ as a possible
generalization of the well-known KS ansatz. Unlike the case $\K = 0$ corresponding
to GKS spacetimes, the KS vector $\bk$ may not be geodetic neither for xKS
spacetimes with aligned matter fields in the context of general relativity nor
even for Einstein xKS spacetimes unless a special relation between the vectors
$\bk$ and $\bm$ appearing in the xKS metric holds. It turns out that such a relation
is compatible with the optical constraint and leads to further simplification of
the Ricci and Riemann tensors of the xKS metric. Unlike the GKS case, xKS spacetimes
with a geodetic $\bk$ are of Weyl type I and thus, in general, may not be
algebraically special.

For the simplest subclass, namely the class of Kundt xKS spacetimes, we have been
able to express the conditions for more special Weyl types determined by the
specific $r$-dependence of the function $\K$ and the form of the Ricci tensor.
It turns out that VSI metrics including type III Ricci-flat \pp waves take the
xKS form with flat background and thus a wide class of explicit examples of Kundt
xKS spacetimes is known. Furthermore, it is shown that type II Ricci-flat \pp waves
belong to the class of xKS spacetimes if they are CSI. 

An example of expanding xKS spacetime, namely the CCLP solution representing
a charged rotating black hole in five-dimensional minimal gauged supergravity,
is also briefly discussed. We have established a null frame, expressed the optical
matrix, and shown that the CCLP black hole is of Weyl type I$_i$. Interestingly,
although the CCLP spacetime is not Einstein nor algebraic special, the corresponding
optical matrix satisfies the optical constraint which, in fact, holds for any
five-dimensional algebraically special Einstein spacetime \cite{OrtaggioPravdaPravdova2012}.

We believe that the xKS form may lead to the discovery of new solutions of general
relativity in higher dimensions in vacuum and also in the presence of matter fields
aligned with the KS vector $\bk$, such as an aligned Maxwell field. Using the xKS
ansatz, one could also obtain new vacuum solutions in more general theories of
gravity, for instance, in the Gauss--Bonnet theory or Lovelock gravities of higher
order. We hope that the results of this paper will be useful for finding such
new solutions in a subsequent work.

%% file: xKS_Ricci_and_Riemann.tex
\section{Frame components of the Ricci and Riemann tensors}
\label{sec:appendix:RicciRiemann}

In this appendix section, we explicitly provide all components of the Ricci and
Riemann tensors for xKS spacetimes \eqref{xKS:metric} with respect to the frame
\eqref{intro:frame:constraints}. The frame vectors $\bl$ and $\bm^{(2)}$ are
identified with the vectors occurring in the xKS metric as $\bk \equiv \bl$ and
$\bm \equiv \bm^{(2)}$, respectively. The null Kerr--Schild vector $\bk$ is assumed
to be geodetic and affinely parametrized.

\subsection{Ricci tensor}

\begin{eqnarray}
  \fl
  R_{00} = 0, \label{xKS:R00} \\
  \fl
  R_{0i} = - \half (\D^2 \K + L_{jj} \D\K + 2 \omega^2 \K ) \delta_{2 i}
    - \half (L_{jj} + \D) (\K \Xi_i)
    - \half \Xi_i \D\K
    - 2 \K A_{2j} S_{ji} \nonumber \\
  \fl \qquad
    - \K L_{2j} L_{ji}
    + \K L_{2i} L_{jj}
    + \half \K (S_{ij} - 3 A_{ij} + \M{i}{j0}) \Xi_j
    + \D\K (L_{2i} + A_{2i}), 
  \label{xKS:R0i} \\
  \fl
  R_{01} = - \D^2\H
    - L_{ii} \D\H
    - 2 \H \omega^2
    - \half \delta_2 \D\K
    - A_{2i} \delta_i \K
    + \half (L_{ii} + \D) (\K (\D\K + \K L_{22})) \nonumber \\
  \fl \qquad
    - \half (2 \K A_{2i} + \M{i}{jj} + \delta_i) (\K \Xi_i)
    + 2 N_{20} \D\K
    + \K \D N_{20}
    - \half M_{ii} \D\K \nonumber \\
  \fl \qquad
    + \K \left( L_{i1} S_{2i} - L_{1i} A_{2i} - A_{ij} M_{ij} + L_{ii} N_{20} + M_{i0} N_{i0} \right)
    + \frac{2 \Lambda}{n-2},
  \label{xKS:R01} \\
  \fl
  R_{ij} = - 2 S_{ij} \D\H
    + 2 \H L_{ik} L_{jk}
    - 2 \H L_{kk} S_{ij}
    + \delta_{(i|} (\K (2 A_{2k} - \Xi_k)) \delta_{k|j)}
    - S_{ij} \delta_2 \K
    \nonumber \\
  \fl \qquad
    + (\K S_{ij} - M_{(ij)}) \D\K
    + \K \Big( (2 L_{[21]} + 2 N_{20} - M_{kk}) S_{ij}
    + 2 M_{(i|0} L_{|j)1}
    + 2 L_{(i|k} M_{|j)k}
    \nonumber \\
  \fl \qquad
    - L_{kk} M_{(ij)}
    + (2 A_{2k} - \Xi_k) \M{k}{(ij)} \Big)
    + \K^2 \Big( (L_{22} + L_{kk}) S_{ij}
    - S_{ik} S_{jk}
    - A_{ik} A_{jk}
    \nonumber \\
  \fl \qquad
    + 2 S_{2(i} A_{j)2}
    + \half \Xi_i \Xi_j
    - L_{2(i} \Xi_{j)} \Big)
    - \Bigg[ \delta_k \D\K
    - 2 L_{k1} \D\K
    - 2 L_{kl} \delta_l \K
    + \K (L_{k2} - \Xi_k) \D\K
    \nonumber \\
  \fl \qquad
    + L_{ll} \delta_k \K
    - 2 \K \delta_l A_{kl}
    - 2 \K \Big( 2 L_{[1l]} S_{kl}
    - 2 L_{l1} A_{kl}
    - L_{[1k]} L_{ll}
    + A_{kl} \M{l}{mm} - A_{lm} \M{k}{lm} \Big)
    \nonumber \\
  \fl \qquad
    - 2 \K^2 \Big( 2 L_{l[l} A_{k]2} L_{ll} - A_{kl} \Xi_l \Big) \Bigg] \delta_{2(i} \delta_{j)k}
    + \Bigg[ \half (\D\K)^2 - \K^2 \omega^2 \Bigg] \delta_{2i} \delta_{2j}
    + \frac{2 \Lambda}{n-2} \delta_{ij},
    \label{xKS:Rij} \\
  \fl
  R_{1i} = - \bigg( 2 L_{[1i]} + \delta_i \bigg) \D\H
    + 2 L_{ij} \delta_j \H
    - L_{jj} \delta_i \H
    + 2 \H \bigg( \bigg( \M{j}{kk} + \delta_j \bigg) A_{ij}
    + 3 L_{[1j]} L_{ij}
    \nonumber \\
  \fl \qquad
    + L_{(1j)} L_{ji}
    - L_{1i} L_{jj}
    + A_{jk} \M{j}{ik}
    \bigg)
    + \half \delta_i \delta_2 \K
    - \delta_i \delta_2 \K
    + \half \D\K \delta_i \K
    - \half \bigg( S_{2i} - \Xi_i \bigg) \Delta \K
    \nonumber \\
  \fl \qquad
    + \half \bigg( N_{2i} + M_{i1} + \K L_{1i} - \K^2 \bigg( L_{2i} - \half \Xi_i \bigg) \bigg) \D\K
    - \bigg( L_{(1i)} - \half \K \bigg( L_{i2} - \Xi_i \bigg) \bigg) \delta_2 \K
    \nonumber \\
  \fl \qquad
    - \bigg( L_{[12]} - N_{20} + \half M_{jj} - \half \K \bigg( L_{22} + L_{jj} \bigg) \bigg) \delta_i \K
    + \bigg( M_{ij} - \half \K L_{ij} \bigg) \delta_j \K
    \nonumber \\
  \fl \qquad
    + \bigg( \H \D\K - \K \D\H \bigg) L_{2i}
    + \K \Bigg[ \delta_2 L_{[1i]}
    + \bigg( \K \bigg( L_{2i} - 2 \Xi_i \bigg) - 2 \delta_i \bigg) L_{[12]}
    + \half \Delta \Xi_i
    + \delta_j M_{[ij]}
    \nonumber \\
  \fl \qquad
    + 2 \H \bigg( 2 L_{j[i} A_{j]2} + L_{j[i} \Xi_{j]} \bigg)
    - 2 L_{[1|1} L_{|2]i}
    - \bigg( 2 L_{[12]} + M_{jj} \bigg) L_{(1i)}
    + \bigg( 2 M_{ij} + \M{j}{i2} \bigg) L_{[1j]}
    \nonumber \\
  \fl \qquad
    + L_{(1j)} M_{ji}
    - \bigg( 2 L_{[1i]} + \K L_{2i} + \delta_i \bigg) \Theta
    - \bigg( 2 L_{ij} - L_{ji} \bigg) M_{j1}
    + 2 L_{ij} N_{[j2]}
    - 2 S_{ij} N_{(j2)}
    \nonumber \\
  \fl \qquad
    + L_{jj} N_{2i}
    - \bigg( N_{[ij]} + \half \M{i}{j1} \bigg) \Xi_j
    + \M{j}{ik} M_{[jk]}
    + \M{j}{kk} M_{[ij]}
    - 2 \bigg( N_{ji} - \K M_{[ij]} \bigg) A_{2j}
    \nonumber \\
  \fl \qquad
    + N_{j(i} \Xi_{j)}
    - \half \K \bigg( L_{12} + \half \K L_{22} \bigg) \bigg( 2 A_{2i} - \Xi_i \bigg)
    + \K L_{1[i} L_{j]j}
    - \K L_{j[i} L_{j]1}
    + \K L_{[i|j} M_{|j]2}
    \nonumber \\
  \fl \qquad
    + \K M_{j[i} \Xi_{j]}
    + \half \K L_{1i} \Xi_{2}
    - \K^2 S_{j[i} \Xi_{j]} \Bigg]
    + \Bigg[ \half \Delta \D\K
    - \half \bigg( \D\K + \K S_{jj} \bigg) \delta_2 \K
    \nonumber \\
  \fl \qquad
    + \half \bigg( 2 L_{[1j]} + \M{j}{kk} + \delta_j \bigg) \bigg( \delta_j \K + 2 \K L_{[1j]} \bigg)
    + \half \bigg( \half \K \bigg( \D\K + \K L_{22} \bigg)
    + N_{jj}
    \nonumber \\
  \fl \qquad
    - \K \bigg( L_{12} - 2 L_{21} + M_{jj} \bigg)
    - \bigg( 2 \H - \K^2 \bigg) S_{jj} \bigg) \D\K
    - \K \bigg( 2 \H A_{jk} + N_{jk} + \K M_{jk} \bigg) A_{jk}
    \nonumber \\
  \fl \qquad
    - \K^2 \bigg( L_{[12]} S_{jj} + A_{2j} \bigg( L_{j1} + \half \K \Xi_j \bigg) \bigg)
    + \frac{2 \K \Lambda}{n - 1}
    \Bigg] \delta_{2i},
    \label{xKS:R1i} \\
  \fl
  R_{11} = \delta_i \delta_i \H
    + N_{ii} \D\H
    + \bigg(4 L_{1i} - 2 L_{i1} + \M{i}{jj} \bigg) \delta_i \H
    - S_{ii} \Delta \H
    + 2 \H \Bigg[ \delta_i L_{1i} - \Delta S_{ii}
    \nonumber \\
  \fl \qquad
    + \bigg( 4 L_{[1i]} + \M{i}{jj} \bigg) L_{1i} - L_{ij} N_{ij} - 2 \M{i}{j1} S_{ij} \Bigg]
    - \half \delta_{\itilde} ( \K \delta_{\itilde} \K )
    - \Delta \delta_2 \K
    + \bigg( \delta_2 \H - N_{21}
    \nonumber \\
  \fl \qquad
    + 4 \H L_{[12]} + \H M_{ii} - \half \K (N_{22} + N_{ii}) + \frac{1}{4} \K^2 M_{22} \bigg) \D\K
    + \K \bigg( \delta_i \H + 4 \H L_{[1i]} \bigg) M_{i0}
    \nonumber \\
%%%%
  \fl \qquad
    + \bigg( \H - \half \K^2 \bigg) \bigg( \delta_2 \D\K + \delta_i (\K \Xi_i) - 2 \delta_i (\K A_{2i})
    + \K \M{i}{jj} \Xi_i + \K^2 L_{2i} L_{2i} - \K^2 L_{22} L_{ii} \bigg)
    \nonumber \\
  \fl \qquad
    - \bigg(\delta_2 \K + 2 \K L_{[12]} + \K M_{ii} \bigg) \D\H
    + \K \Delta (\K S_{ii})
    + \half (\K \delta_2 - 2 \Delta)(\K M_{ii})
    - 2 \Delta (\K L_{[12]})
    \nonumber \\
  \fl \qquad
    - \bigg( L_{11} - 2 \K L_{12} + \frac{3}{2} \K L_{21} - \K N_{20} - \frac{1}{4} \K^2 L_{22} + \half \K^2 S_{ii} \bigg) \delta_2 \K
    - \bigg( 2 N_{2i} + 2 \K L_{1i}
    \nonumber \\
  \fl \qquad
    - \half \K L_{i1} + \half \K \M{i}{jj} - \K M_{i2} - \K^2 S_{2i}
    + \frac{3}{4} \K^2 \Xi_i \bigg) \delta_i \K
    - \K^2 \bigg( \M{i}{jj} - \half \K \Xi_i + \delta_i \bigg) L_{(1i)}
    \nonumber \\
  \fl \qquad
    - \K \bigg (3 L_{1i} - 2 L_{i1} + \M{i}{jj} + \K M_{i0} - \delta_i \bigg) N_{2i}
    + \K \bigg( \H \bigg( 2 L_{i1} - 2 \M{i}{jj} + \K \Xi_i \bigg) - \K M_{i1}
    \nonumber \\
  \fl \qquad
    - \K^2 \bigg( 2 L_{i1} + M_{i2} - \M{i}{jj} \bigg)
    + \K^3 \bigg( A_{2i} - \Xi_i \bigg) \bigg) A_{2i}
    - \K \bigg( 2 L_{11} - \K \bigg(2 L_{12} - L_{21} + M_{ii}
    \nonumber \\
  \fl \qquad
    + 2 N_{20} + \K \bigg( \half L_{22} - S_{ii} \bigg) + \delta_2 \bigg) \bigg) L_{[12]}
    + \K \bigg( 2 N_{i1} - \K N_{i2} + 2 \K^2 L_{[1i]} \bigg) \bigg( S_{2i} - \half \Xi_i \bigg)
    \nonumber \\
  \fl \qquad
    - \K \bigg ( L_{11} - \K N_{20} \bigg) M_{ii}
    - \K \bigg( 2 \M{i}{j1} - \K \M{i}{j2} \bigg) M_{(ij)}
    - \K N_{20} N_{ii}
    - \K M_{ij} N_{ij}
    \nonumber \\
  \fl \qquad
    - \half \K^2 \bigg(2 L_{1i} - L_{i1} - 2 M_{i2} + \K \Xi_i \bigg) L_{1i}
    + \half \K^2 \bigg( L_{i1} - M_{i2} \bigg) L_{i1}
    + \half \K^2 M_{ij} M_{ji}
    \nonumber \\
  \fl \qquad
    + \K^2 \bigg( N_{ij} + 2 \M{i}{j1} \bigg) S_{ij}
    + \half \K^2 N_{22} S_{ii}
    + \frac{1}{4} \K^3 M_{i2} \Xi_i
    - \K^3 A_{ij} M_{ij}
    - \half \K^4 A_{ij} A_{ij}
    \nonumber \\
  \fl \qquad
    + \frac{(n-3) \Lambda}{(n-1)(n-2)} \K^2,
    \label{xKS:R11}
\end{eqnarray}
where $\Xi_i = L_{2i} + M_{i0}$ and $\Theta = L_{21} - N_{20}$.

\subsection{Riemann tensor}

\begin{eqnarray}
  \fl
  R_{0i0j} = 0,
  \label{xKS:R0i0j} \\
  \fl
  R_{010i} = \half \D^2\K \, \delta_{2i}
    + \half \D ( \K \Xi_i )
    + \half M_{i0} \D\K
    - \half \K \Xi_j \M{i}{j0},
  \label{xKS:R010i} \\
  \fl
  R_{0ijk} = \bigg( L_{i[j} \delta_{k]2}
    - A_{jk} \delta_{2i} \bigg) \D\K
    + \K \bigg( 2 L_{2[j} S_{k]i}
    + L_{i[j} \Xi_{k]}
    - A_{jk} \Xi_i
    - 2 A_{il} L_{l[j} \delta_{k]2}
    \nonumber \\
  \fl \qquad
    - 2 S_{l[j} A_{k]l} \delta_{2i} \bigg),
  \label{xKS:R0ijk} \\
  \fl
  R_{0101} = \D^2\H
    - \frac{1}{4} (\D\K)^2
    - \bigg( \half \K L_{2 2}    
    + N_{20} \bigg) \D\K
    + \D (\K \Theta)
    - \K \Xi_i N_{i0}
    - \frac{1}{4} \K^2 \Xi_i \Xi_i
    \nonumber \\
  \fl \qquad
    - \frac{2 \Lambda}{(n - 2)(n - 1)}, \\
  \fl
  R_{01ij} = - 2 A_{ij} \D\H
    - 4 \H S_{k[i} A_{j]k}
    + \delta_{[i}\D\K \, \delta_{j]2}
    + \bigg( \K L_{2[i} \delta_{j]2}
    - M_{[ij]} \bigg) \D\K
    - M_{[i|0} \, \delta_{j]}\K
    \nonumber \\
  \fl \qquad
    - ( \delta_k\K + 2 \K L_{[1k]} ) L_{k[i} \delta_{j]2}
    + \K \bigg( L_{2[i} L_{1|j]}
    + L_{2[i} L_{j]1}
    - 2 A_{ij} \Theta
    - L_{k[i} M_{j]k}
    - \Xi_k \M{k}{[ij]}
    \nonumber \\
  \fl \qquad
    + L_{k[i} M_{k|j]}
    + \delta_k \Xi_l \delta_{k[i} \delta_{j]l} \bigg)
    + \K^2 L_{2[i} \Xi_{j]}, \\
  \fl
  R_{0i1j} = - L_{ij} \D\H
    + 2 \H A_{ik} L_{kj}
    + \half (\K S_{ij} - M_{ij}) \D\K
    - \half L_{2j} \delta_i \K
    - \half \delta_j (\K \Xi_i)
    \nonumber \\
  \fl \qquad
    + \K \bigg( S_{ij} N_{20}
    - \Theta A_{ij}
    - L_{(1i)} L_{2j}
    + \Xi_{(i} L_{j)1}
    + L_{kj} M_{[ik]}
    - \half \Xi_k \M{k}{ij} \bigg)
    + \half \K^2 \bigg( L_{22} S_{ij}
    \nonumber \\
  \fl \qquad
    - \Xi_i L_{2j}
    + \half \Xi_i \Xi_j \bigg)
    - \half \Bigg[ \delta_j \D\K
    + \K L_{2j} \D\K
    - L_{kj} \delta_k \K
    - 2 \K L_{[1k]} L_{kj} \Bigg] \delta_{2i}
    \nonumber \\
  \fl \qquad
    + \Bigg[ \bigg( L_{l1}
    + \half \K \Xi_l
    + \K A_{2l} \bigg) \D\K
    - 2 \K L_{k1} A_{lk}
    - \K^2 A_{lk} \Xi_k \Bigg] \delta_{2(i} \delta_{j)l}
    + \frac{1}{4} (\D\K)^2 \delta_{2i} \delta_{2j} \nonumber \\
  \fl \qquad
    + \frac{2 \Lambda}{(n - 2)(n - 1)} \delta_{ij},
  \label{xKS:R0i1j} \\
  \fl
  R_{ijkl} = 4 \H \bigg( A_{ij} A_{kl} + A_{l[i} A_{j]k} + S_{l[i} S_{j]k} \bigg)
    + 2 \K \bigg( A_{ij} M_{[kl]}
    + A_{kl} M_{[ij]}
    + A_{l[i} M^{\mathrm{A}}_{j]k}
    \nonumber \\
  \fl \qquad
    + M^{\mathrm{A}}_{l[i} A_{j]k}
    + S_{l[i} M^{\mathrm{S}}_{j]k}
    + M^{\mathrm{S}}_{l[i} S_{j]k} \bigg)
    + 2 \K^2 S_{k[i} S_{j]l}
    - 4 \K^2 \delta_{2[i} A_{j]s} A_{s[k} \delta_{l]2}
    \nonumber \\
  \fl \qquad
    + 2 \Bigg[
    L_{[n|m} \delta_{|p]} \K
    + A_{np} \delta_m \K
    + \K \bigg( \delta_m A_{np}
    - 2 A_{s[n} \M{s}{p]m}
    - L_{[1n]} L_{pm}
    + L_{[1p]} L_{nm} \bigg)
    \nonumber \\
  \fl \qquad
    + 2 \K^2 A_{2[n} S_{p]m}
    \Bigg] \bigg( \delta_{2[i} \delta_{j]m} \delta_{nk} \delta_{pl} + 
    \delta_{2[k} \delta_{l]m} \delta_{ni} \delta_{pj} \bigg)
    + \frac{4\Lambda}{(n-2)(n-1)} \delta_{i[k} \delta_{l]j},
    \label{xKS:Rijkl} \\
  \fl
  R_{101i} = \delta_i \D\H
    + 2 L_{[1i]} \D\H
    - L_{ji} \delta_j \H
    + 2 \H ( L_{j1} S_{ij} - L_{1j} L_{ji})
    - \H ( A_{2i} \D\K - \K A_{ij} \Xi_j)
    \nonumber \\
  \fl \qquad
    + \K L_{2 i} \D\H
    - \bigg( \half M_{i1} + N_{2i} - \frac{1}{4} \K M_{i2} + \frac{1}{4} \Xi_i \K^2 \bigg) \D\K
    + \delta_i (\K \Theta)
    - \half \Delta (\K \Xi_i)
    \nonumber \\
  \fl \qquad
    + L_{2i} \Delta \K
    - \frac{1}{4} \bigg( \D\K + 2 L_{21} + \K L_{22} \bigg) \bigg( \delta_i \K + 2 \K L_{(1i)} \bigg)
    + \K \bigg( 2 L_{[1i]} + \K L_{2i} \bigg) \Theta 
    \nonumber \\
  \fl \qquad
    + \K \bigg( L_{11} L_{2i}
    + L_{j1} M_{[ij]}
    + L_{ji} M_{j1}
    + L_{ji} N_{2j} \bigg)
    + \half \K \bigg( \M{i}{j1} + \K M_{[ij]} - 2 N_{ji} \bigg) \Xi_j
    \nonumber \\
  \fl \qquad
    - \frac{1}{4} \K^2 \bigg( 2 L_{21} + \K L_{22} \bigg) \Xi_i
    - \half \Bigg[ \Delta DK
    + \half \K (\D\K)^2
    - \half \D\K \delta_2 \K
    - \half \K \bigg( L_{12} - 3 L_{21}
    \nonumber \\
  \fl \qquad
    - \K L_{22} \bigg) \D\K
    - \bigg( L_{j1} + \half \K \Xi_j \bigg) \bigg( \delta_j \K + 2 \K L_{[1j]} \bigg)
    + \frac{ 4 \K \Lambda}{(n - 1)(n - 2)} \Bigg] \delta_{2i}, \\
  \fl
  R_{1ijk} = 2 L_{[j|i} \delta_{|k]} \H
    + 2 \delta_i (\H A_{jk})
    + 4 \H \bigg( L_{[1i]} A_{jk} + A_{i[j} L_{k]1} - L_{1[j} L_{k]i}
    - A_{l[j} \M{l}{k]i} \bigg)
    \nonumber \\
  \fl \qquad
    + 2 \H \K \bigg( S_{i[j} \Xi_{k]} + 2 A_{2[j} S_{k]i} \bigg)
    + M_{[j|i} \delta_{|k]} \K
    - \K S_{i[j} \delta_{k]} \K
    + M_{[jk]} \delta_i \K
    + \K \bigg( \delta_i M_{[jk]}
    \nonumber \\
  \fl \qquad
    + 2 L_{[1i]} M_{[jk]}
    - L_{1[j} M_{k]i}
    - L_{[j|1} M_{i|k]}
    + 2 A_{i[j} M_{k]1}
    - A_{jk} M_{i1}
    + 2 N_{2[j} L_{k]i}
    - A_{jk} N_{2i}
    \nonumber \\
  \fl \qquad
    - 2 M^{\mathrm{A}}_{l[j} \M{l}{k]i}
    - N_{[j|i} \Xi_{|k]} \bigg)
    - \K^2 \bigg( \delta_{[j} A_{k]i}
    + \half \delta_i A_{jk}
    - L_{1[j} L_{k]i}
    - L_{[j|1} L_{i|k]}
    - L_{[1i]} A_{jk}
    \nonumber \\
  \fl \qquad
    + A_{il} \M{l}{[jk]}
    - A_{jl} \M{l}{[ik]}
    + A_{kl} \M{l}{[ij]}
    - S_{i[j} M_{k]2}
    - M^{\mathrm{S}}_{i[j} \Xi_{k]} \bigg)
    - \K^3 S_{i [j} \Xi_{k]}
    - \Bigg[ \delta_{[j} \delta_{k]} \K
    \nonumber \\
  \fl \qquad
    - N_{[jk]} \D\K
    + \bigg( \K A_{2l} + \half \K \Xi_l + L_{l1} \bigg) \delta_{l[j} \delta_{k]} \K
    - \M{l}{[jk]} \delta_l \K
    + \K \bigg( \delta_j L_{[1k]}
    - \delta_k L_{[1j]}
    \nonumber \\
  \fl \qquad
    - L_{1[j} L_{k]1}
    - 2 L_{[1l]} \M{l}{[jk]}
    + 2 N_{l[j} A_{k]l} \bigg)
    + \K^2 \bigg( 2 A_{2[j} L_{k]1}
    - \bigg( M_{lm} + \M{m}{l2} \bigg) \delta_{l[j} A_{k]m}
    \nonumber \\
  \fl \qquad
    + \bigg( \M{l}{[jk]} + \delta_{l[j} \delta_{k]} \bigg) A_{2l}
    - \bigg( L_{[12]} + \half \delta_2 \bigg) A_{jk}
    + L_{[1l]} \Xi_{[j} \delta_{k]l} \bigg)
    + \K^3 \bigg( A_{2[j} A_{k]2}
    + A_{2[j} \Xi_{k]}
    \nonumber \\
  \fl \qquad
    - S_{2[j} S_{k]2} \bigg)
    \Bigg] \delta_{2i}
    + \Bigg[ \delta_l \delta_i \K
    + \bigg( (\K^2 - 2 \H) S_{il}
    + N_{il} - \K M_{(il)} \bigg) \D\K
    - \K S_{il} \delta_2 \K
    \nonumber \\
  \fl \qquad
    + 2 \bigg( L_{[1i]} - \half \K A_{2i} \bigg) \delta_l \K
    + \bigg( L_{1l} + \half \K \Xi_l \bigg) \delta_i \K
    + \M{m}{il} \delta_m \K
    + 2 \K \bigg( \delta_l L_{[1i]}
    + L_{[1i]} L_{1l}
    \nonumber \\
  \fl \qquad
    + L_{[1m]} \M{m}{il}
    + \bigg( 2 \H A_{ml} + N_{ml} \bigg) A_{im} \bigg)
    - 2 \K^2 \bigg( \delta_{[i} A_{l]2}
    + \half \delta_2 A_{ik}
    + \bigg( L_{l1} - \half \K \Xi_l \bigg) A_{2i}
    \nonumber \\
  \fl \qquad
    + L_{[12]} L_{li}
    + \half \bigg( M_{lm} + \M{m}{l2} \bigg) A_{im}
    + A_{lm} \M{m}{[2i]}
    + A_{2m} \M{m}{[il]}
    + \bigg( A_{2l} - \half \Xi_l \bigg) L_{[1i]}
    \bigg)
    \nonumber \\
  \fl \qquad
    - \half \bigg( \bigg( \delta_l \K + 2 \K L_{[1l]} - 2 \K^2 A_{2l} \bigg) \D\K
    - 2 \K \bigg( \delta_m \K
    + 2 \K L_{[1m]} \bigg) A_{lm}
    \bigg) \delta_{2i}
    \nonumber \\
  \fl \qquad
    + \frac{4 \K \Lambda}{(n - 2)(n - 1)} \delta_{il}
    \Bigg]\delta_{2[j} \delta_{k]l}, \\
  \fl
  R_{1i1j} = \delta_{(i} \delta_{j)} \H
    + N_{(ij)} \D\H
    - S_{ij} \Delta \H
    + 4 L_{1(i} \delta_{j)} \H
    - 2 L_{(i|1} \delta_{|j)} \H
    + \M{k}{(ij)} \delta_k \H
    \nonumber \\
  \fl \qquad
    - \H \Bigg[ 2 \Delta S_{ij}
    - 2 \delta_{(i|} L_{1|j)}
    - 4 L_{1i} L_{1j}
    + 4 L_{1(i} L_{j)1}
    - 2 L_{1k} \M{k}{(ij)}
    - N_{k(i} A_{j)k}
    + 2 N_{k(i} S_{j)k}
    \nonumber \\
  \fl \qquad
    + 4 S_{k(i} \M{k}{j)1}
    - 2 \H A_{k(i} S_{j)k} \Bigg]
    - \K \bigg( 3 A_{2k} + S_{2k} \bigg) \delta_{k(i} \delta_{j)} \H
    - \half \K \delta_{(i} \delta_{j)} \K
    - \frac{1}{4} \delta_i \K \delta_j \K
    \nonumber \\
  \fl \qquad
    - \half \K N_{(ij)} \D\K
    + \K \Delta (\K S_{ij})
    - \Delta (\K M_{(ij)})
    + (\H \D\K - \K \D\H) M_{(ij)}
    - \Bigg[ M_{k1} + 2 N_{2k}
    \nonumber \\
  \fl \qquad
    - \H L_{k2} - \K \bigg( M_{k2} - \frac{3}{2} L_{1k} \bigg)
    - \half \K^2 \bigg( 3 A_{2k} + S_{2k} - \frac{5}{2} \Xi_{k} \bigg) \Bigg] \delta_{k(i} \delta_{j)} \K
    + \Xi_{(i} \delta_{j)} (\H \K)
    \nonumber \\
  \fl \qquad
    - \half \K^2 S_{ij} \delta_2 \K
    - \half \K \M{\tilde{k}}{(ij)} \delta_{\tilde{k}} \K
    - \K \Bigg[
    \delta_{(i|} N_{2|j)}
    + L_{11} M_{(ij)}
    + N_{20} N_{(ij)}
    + M_{k(i|} N_{k|j)}
    \nonumber \\
  \fl \qquad
    + \bigg( \Xi_k - 2 L_{2k} \bigg) \delta_{k(i} N_{j)1}
    + L_{[1k]} \bigg( 2 N_{2l} + M_{l1} \bigg) \delta_{k(i} \delta_{j)l}
    + \bigg( L_{1(i} \delta_{j)k} + \M{k}{(ij)} \bigg) N_{2k}
    \nonumber \\
  \fl \qquad
    + 2 M^{\mathrm{S}}_{k(i} \M{k}{j)|1}
    \Bigg]
    - \K^2 \Bigg[ \bigg( \delta_{k} L_{(1l)}
    - \bigg( M_{k2} - \frac{3}{2} L_{(1k)} \bigg) L_{[1l]} \bigg) \delta_{k(i} \delta_{j)l}
    + \bigg( \Theta - \half \delta_2 \bigg) M_{(ij)}
    \nonumber \\
  \fl \qquad
    - A_{2(i} \delta_{j)k} \bigg( M_{k1} - 2 N_{2k} \bigg)
    - 2 L_{2(i} \delta_{j)k} N_{[2k]}
    + L_{(1\tilde{k})} \M{\tilde{k}}{(ij)}
    - \half M^{\mathrm{S}}_{ik} M_{kj}
    + \half M^{\mathrm{A}}_{ik} M_{jk}
    \nonumber \\
  \fl \qquad
    - \bigg( N_{kl} + 2 \M{k}{l1} \bigg) S_{k(i} \delta_{j)l}
    - \M{k}{(i|2} M^{\mathrm{S}}_{|j)k}
    - \half N_{22} S_{ij}
    + \bigg( N_{2k} - \half N_{k2} \bigg) \Xi_{(i} \delta_{j)k}
    \Bigg]
    \nonumber \\    
  \fl \qquad
    - \K^3 \Bigg[ \delta_{(i} A_{i)2}
    + \half \delta_{(i} \Xi_{j)}
    + \bigg( L_{[12]} + \half M_{22} \bigg) S_{ij}
    - S_{k(i} M^{\mathrm{A}}_{j)k}
    + 3 A_{2(i} M^{\mathrm{A}}_{j)2}
    - \half M_{k(i} A_{j)k}
    \nonumber \\
  \fl \qquad
    - \half \bigg( L_{2k} - \frac{3}{2} \Xi_k \bigg) L_{[1l]} \delta_{k(i} \delta_{j)l}
    - \bigg( A_{2k} - \half \Xi_k \bigg) \M{k}{(ij)}
  \Bigg]
    - \K^4 \Bigg[ A_{2(i} S_{j)2}
    + \half A_{ik} A_{jk}
    \nonumber \\
  \fl \qquad
    - \half S_{22} S_{ij}
    + \half S_{2i} S_{2j} \Bigg]
    - \H \K \Bigg[ \bigg( 2 L_{1k} + \delta_k \bigg) A_{2(i} \delta_{j)k}
    - \bigg( 3 A_{2k} - S_{2k} \bigg) L_{[1l]} \delta_{k(i} \delta_{j)l}
    \nonumber \\
  \fl \qquad
    - (2 L_{[1k]} + \delta_{k}) \Xi_{(i} \delta_{j)k}
    - (\Xi_k - 2 A_{2k}) \M{k}{(ij)}
    - 2 A_{k(i} M^{\mathrm{S}}_{j)k}
    + 2 S_{k(i} M^{\mathrm{A}}_{j)k}
    - \K \bigg(
    3 A_{2i} A_{2j}
    \nonumber \\
  \fl \qquad
    + S_{2i} S_{2j}
    - S_{22} S_{ij}
    - A_{2(i} \Xi_{j)}
    + 2 S_{\tilde{k}(i} A_{j)\tilde{k}}
    \bigg) \Bigg]
    - \Bigg[ \Delta \delta_k \K
    + \D\H \delta_k \K
    - \D\K \delta_k \H
    \nonumber \\
  \fl \qquad
    - 2 \K A_{kl} \delta_l \H
    - \bigg( \H - \half \K^2 \bigg) \delta_k \D\K
    + \frac{1}{4} \K \D\K \delta_k \K
    + 2 \K \bigg( L_{[1k]} - \K A_{2k} \bigg) \D\H
    \nonumber \\
  \fl \qquad
    - \frac{1}{4} \delta_k \delta_2 \K^2
    + \bigg( N_{k1}
    - 4 \H L_{[1k]}
    + \half \K N_{2k}
    + 3 \H \K A_{2k}
    + \half \K^2 L_{[1k]}
    - \half \K^2 M_{(2k)} \bigg) \D\K
    \nonumber \\
  \fl \qquad
    + \bigg( \M{l}{k1} + \bigg( \H - \half \K^2 \bigg) L_{kl} + \half \K M_{kl} + \frac{3}{2} \K^2 A_{kl} \bigg) \delta_l \K
    + \bigg( L_{11} + \K \Theta - \K L_{12} \bigg) \delta_k \K
    \nonumber \\
  \fl \qquad
    - \K \bigg( L_{1k} - \half L_{k1} - \frac{5}{2} \K A_{2k} + \frac{1}{4} \K \Xi_k \bigg) \delta_2 \K
    + 2 \bigg( L_{11} - \K N_{20} + \Delta \bigg) \bigg( \K L_{[1k]} - \K^2 A_{2k} \bigg)
    \nonumber \\
  \fl \qquad
    - \K^2 \bigg( 2 L_{1k} - L_{k1} - 3 \K A_{2k} + \half \K \Xi_k + \delta_k \bigg) L_{[12]}
    - \K^2 \bigg( N_{lk} + 2 \M{l}{k1} - \K M_{[lk]}
    \nonumber \\
  \fl \qquad
    - \K \M{l}{k2} - \K \delta_{kl} \delta_2 \bigg) A_{2l} + 2 \K \bigg( N_{l1} + \K N_{2l} - 2 \H L_{1l}
    + \K \bigg( \H - \half \K^2 \bigg) \bigg( A_{2l} - \Xi_l \bigg)
    \nonumber \\
  \fl \qquad
    + \K^2 L_{(1l)} \bigg) A_{kl}  
    + \K \bigg( 2 \M{l}{k1} + \K M_{kl} + 2 \H L_{kl} - \K^2 S_{kl} \bigg) L_{[1l]}
    \Bigg] \delta_{2(i} \delta_{j)k}
    \nonumber \\    
%%%%
  \fl \qquad
    + \frac{1}{4} \Bigg[ \K \bigg( \delta_2 \K + 2 \K^2 L_{[12]} \bigg) \D\K
    - \bigg( \delta_k \K + 2 \K L_{[1k]} - \K^2 A_{2k} \bigg)^2
    - \K^4 A_{2k} A_{2k}
    \Bigg] \delta_{2i} \delta_{2j}
    \nonumber \\    
%%%%%
  \fl \qquad
    + \frac{\K^2 \Lambda}{(n-2)(n-1)} (\delta_{ij} - \delta_{2i} \delta_{2j}),
\end{eqnarray}
where $M^{\mathrm{S}}_{ij} = M_{(ij)}$, $M^{\mathrm{A}}_{ij} = M_{[ij]}$,
$\Xi_i = L_{2i} + M_{i0}$, and $\Theta = L_{21} - N_{20}$.

%% file: xKS_general_kundt.tex
\section{Kundt spacetimes}
\label{sec:appendix:Kundt}

The Kundt class is defined geometrically as spacetimes admitting a congruence
of non-expanding, non-shearing, and non-twisting null geodesics $\bl$, i.e.\
$L_{i0} = L_{ij} = 0$. It is always possible to choose an affine parametrization
so that $L_{10} = 0$. Furthermore, employing the remaining freedom of the choice
of the null frame, we can set $L_{1i} = L_{i1}$ and therefore the covariant
derivative of $\bl$ read \cite{ColeyHervikPapadopoulosPelavas2009}
\begin{equation}
  \ell_{a;b} = L_{11} \ell_a \ell_b + 2 L_{1i} \ell_{(a} m^{(i)}_{b)}.
  \label{Kundt:nabla_l}
\end{equation}
In fact, one may also transform away all $L_{1i}$ except one \cite{PhD}, but it is not
necessary for the following discussion. Note also that if $L_{1i}$ is non-zero,
$L_{11}$ can be always eliminated using null rotations with $\ell$ fixed.

Kundt spacetimes in arbitrary dimension have been recently studied from various
points of view \cite{ColeyHervikPelavas2005,PodolskyZofka2008,
ColeyHervikPapadopoulosPelavas2009,PodolskySvarc2013b,PodolskySvarc2013a} 
and they prove to be important in higher order theories of gravity since the
class of Einstein Kundt spacetimes contains universal metrics \cite{HervikPravdaPravdova2013}
and also some non-Einstein Kundt spacetimes are vacuum solutions of quadratic
gravity \cite{MalekPravda2011}.
Although the algebraic classification of the Kundt class was completely presented
in terms of constraints on the metric functions in \cite{PodolskySvarc2013a},
for certain purposes the following procedure could be more convenient. Without
imposing an explicit form of the metric, Kundt spacetimes can be classified
according to the canonical form of $\ell_{a;b}$, which have a clear geometric
interpretation, and for each subclass the Weyl type can be determined in terms
of $L_{11}$, $L_{1i}$ and the Ricci tensor.

The subclass $L_{1i} = 0$ represents spacetimes with a recurrent null vector
(RNV) field. Obviously, the direction of such a recurrent vector $\bl$ remains
invariant under parallel transport along any curve. The holonomy group of RNV
spacetimes is $Sim(n-2)$ \cite{GibbonsPope2007}. Furthermore, if $L_{11}$ also
vanishes, spacetimes belonging to this class admit a covariantly constant null
vector (CCNV) $\bl$ and are referred to as \pp waves. In this case, $\bl$ is
parallelly transported along any curve and the holonomy group specializes to
Euclidean group $E(n-2)$ or a subgroup thereof.

Using \eqref{Kundt:nabla_l} along with the Ricci identities and their appropriate
contractions, we can directly obtain relevant components of the Ricci tensor
\begin{eqnarray}
  R_{00} = 0, \qquad
  R_{0i} = \D L_{1i} + L_{1k} \M{k}{i0}, \label{Kundt:R0i} \\
  R_{01} = \D L_{11} + \delta_i L_{1i} - 2 L_{1i} N_{i0} + L_{1i} \M{i}{jj} \label{Kundt:R01}
\end{eqnarray}
and the Riemann tensor
\begin{eqnarray}
  R_{0i0j} = R_{0ijk} = 0, \qquad
  R_{010i} = - \D L_{1i} - L_{1j} \M{j}{i0}, \\
  R_{0i1j} = \delta_j L_{1i} + L_{1k} \M{k}{ij} - L_{1j} L_{1i}, \\
  R_{01ij} = - 2 \delta_{[i} L_{1|j]} + 2 L_{1k} \M{k}{[ij]}, \\
  R_{0101} = - \D L_{11} - L_{1i} L_{1i} + 2 L_{1i} N_{i0}, \\
  R_{011i} = \delta_i L_{11} - 2 L_{1j} N_{ji} - \T L_{1i} - L_{1j} \M{j}{i1},
\end{eqnarray}
respectively. Subsequently, it allows us to express also the corresponding
components of the Weyl tensor
\begin{eqnarray}
  C_{0i0j} = 0, \qquad
  C_{0ijk} = \frac{2}{n - 2} (\D L_{1l} + L_{1m} \M{m}{l0}) \delta_{i[j} \delta_{k]l},
  \label{Kundt:C0ijk} \\
  C_{010i} = - \frac{n - 3}{n - 2} (\D L_{1i} + L_{1j} \M{j}{i0}),
  \label{Kundt:C010i} \\
  C_{0i1j} = - \frac{1}{n - 2} ( \D L_{11} + \delta_k L_{1k}
    - 2 L_{1k} N_{k0} + L_{1k} \M{k}{ll}) \delta_{ij} \nonumber \\
    \qquad + \delta_j L_{1i} + L_{1k} \M{k}{ij} - L_{1j} L_{1i}
    - \frac{1}{n - 2} R_{ij} + \frac{1}{(n - 2)(n - 1)} R \delta_{ij}, \\
  C_{01ij} = - 2 \delta_{[i} L_{1|j]} + 2 L_{1k} \M{k}{[ij]}, \\
  C_{0101} = - \frac{n - 4}{n - 2} (\D L_{11} - 2 L_{1i} N_{i0})
    + \frac{2}{n - 2} (\delta_i L_{1i} + L_{1i} \M{i}{jj}) \nonumber \\
    \qquad - L_{1i} L_{1i} - \frac{1}{(n - 2)(n - 1)} R, \\
  C_{011i} = \delta_i L_{11} - 2 L_{1j} N_{ji} - \T L_{1i} - L_{1j} \M{j}{i1} - \frac{1}{n - 2} R_{1i}.
\end{eqnarray}
Now, for each subclass of Kundt spacetimes, we inspect these components of the
Weyl tensor and compare them with the classification scheme based on null alignment
and the spin-type refinement \cite{Coleyetal2004,MilsonColeyPravdaPravdova2004},
see also \cite{OrtaggioPravdaPravdova2012} for recent review.

It is known that Kundt spacetimes are in general of type I with $R_{00} = 0$ and
of type II if and only if $R_{0i} = 0$ \cite{OrtaggioPravdaPravdova2007}. More
precisely, type I Kundt spacetimes are of subtype I(b) since $C_{0ijk} C_{0ijk} =
\frac{2}{n-3} C_{010i} C_{010i}$ as follows from \eqref{Kundt:C0ijk} and
\eqref{Kundt:C010i}. 

For $L_{1i} = 0$, the components of the Ricci tensor \eqref{Kundt:R0i} and
\eqref{Kundt:R01} reduce to $R_{00} = R_{0i} = 0$, $R_{01} = \D L_{11}$ which
also implies that RNV spacetimes are of Weyl type II. The conditions for each
possible subtype are discussed in more detail in table \ref{tab:xKS:Kundt:recurent}.
It turns out that RNV spacetimes are in general of type II(d), while Einstein RNV
spacetimes for which $\D L_{11} = \frac{2\Lambda}{n-2}$ are of type II(bd).
Ricci-flat RNV spacetimes are of type II(abd) with $\D L_{11} = 0$ and conversely
type II(a) Einstein RNV spacetimes are necessarily Ricci-flat. It was pointed out
in \cite{HervikPravdaPravdova2013} that for type III(a) Ricci-flat RNV spacetimes,
$L_{11}$ can be always transformed away by a boost \eqref{intro:frame:boost} where
$\lambda$ subject to $L_{11} = - \lambda^{-1} \T \lambda$ since $\D L_{11} =
\delta_i L_{11} = 0$ and thus such spacetimes are \pp waves. Therefore, the entire
class of type II(a) Einstein RNV spacetimes consists only of Ricci-flat CCNV
spacetimes.

The conditions
for types II(c) and III(b) are not given only in terms of $L_{11}$ and the Ricci
tensor. In order to distinguish these subtypes one has to necessarily know
a particular form of the metric to express $C_{ijkl}$ and $C_{1ijk}$, respectively.
Type II RNV spacetimes are of subtype II(c) if and only if
\begin{eqnarray}
  C_{ijkl} = \frac{4}{(n-3)(n-4)} \Bigg( \D L_{11} - \frac{2n - 5}{(n-1)(n-2)} R \Bigg) \delta_{i[k} \delta_{l]j} \nonumber \\
    \qquad + \frac{4}{(n-2)(n-4)} (\delta_{i[k} R_{l]j} - \delta_{j[k} R_{l]i}).
  \label{Kundt:condition:II(c)}
\end{eqnarray}
The necessary and sufficient condition for type III RNV spacetimes to be of subtype
III(b) is
\begin{equation}
  C_{1ijk} C_{1ijk} = \frac{2}{n-3} \Bigg( \delta_i L_{11} - \frac{1}{n-2} R_{1i} \Bigg) \Bigg( \delta_i L_{11} - \frac{1}{n-2} R_{1i} \Bigg).
  \label{Kundt:condition:III(b)}
\end{equation}

\fulltable{\label{tab:xKS:Kundt:recurent}Algebraic classification of RNV spacetimes.
For each subtype, the necessary and sufficient conditions are presented for the
case of Einstein spacetimes and for general form of the Ricci tensor.}
  \br
  &&\centre{2}{Condition} \\
  \ns
  &&\crule{2} \\
  Type & Subtype & general Ricci tensor & Einstein spacetimes \\
  \mr
  \multirow{4}{*}{II} & II(a) & $\frac{R}{n-1} + (n-4) \D L_{11} = 0$ & $\Lambda = 0$ \\
  & II(b) & $R_{ij} - \frac{R}{n-2} \delta_{ij} + \frac{2 \D L_{11}}{n-2} \delta_{ij} = 0$ & identically satisfied \\
  & II(c) & eq.\ \eqref{Kundt:condition:II(c)} & $C_{ijkl} = \frac{8 \Lambda}{(n-1)(n-2)(n-3)} \delta_{i[k} \delta_{l]j}$ \\
  & II(d) & identically satisfied & identically satisfied \\
  \mr
  \multirow{2}{*}{III} & III(a) & II(abc) $\land$ $\delta_i L_{11} - \frac{1}{n-2} R_{1i} = 0$ & II(ac) $\land$ $\delta_i L_{11} = 0$ \\
  & III(b) & II(abc) $\land$ eq.\ \eqref{Kundt:condition:III(b)} & II(ac) $\land$ $C_{1ijk} C_{1ijk} = \frac{2}{n-3} (\delta_i L_{11})^2$ \\
  \mr
  N & & III(ab) & III(ab) \\
  \br
\endfulltable

In the case of \pp waves, it follows that $R_{00} = R_{0i} = R_{01} = 0$ and the
conditions determining the corresponding subtypes further simplify, see table
\ref{tab:xKS:Kundt:ppwaves}. In general, \pp waves are of type II(d). Since
$R_{01} = 0$, Einstein \pp waves are necessarily Ricci-flat and these are of type
II(abd).

\fulltable{\label{tab:xKS:Kundt:ppwaves}Algebraic classification of \pp waves
(CCNV spacetimes). The necessary and sufficient conditions determining the
corresponding subtype are given for general Ricci tensor and for the Ricci-flat
case.
}
  \br
  && \centre{2}{Condition} \\
  \ns
  && \crule{2} \\
  Type & Subtype & general Ricci tensor & Ricci-flat \\
  \mr
  \multirow{4}{*}{II} & II(a) & $R = 0$ & identically satisfied \\
  & II(b) & $R_{ij} - \frac{R}{n-2} \delta_{ij} = 0$ & identically satisfied \\
  & II(c) & eq.\ \eqref{Kundt:condition:II(c)} with $L_{11} = 0$ & $C_{ijkl} = 0$ \\
  & II(d) & identically satisfied & identically satisfied \\
  \mr
  \multirow{2}{*}{III} & III(a) & II(abc) $\land$ $R_{1i} = 0$ & II(c) \\
  & III(b) & II(abc) $\land$ $C_{1ijk} C_{1ijk} = \frac{2}{(n-2)(n-3)} R_{1i} R_{1i}$ & II(c) $\land$ $C_{1ijk} C_{1ijk} = 0$ \\
  \mr
  N & & III(ab) & III(b) \\
  \br
\endfulltable

Some properties of the above-mentioned subclasses of Kundt spacetimes determined
by the geometry of the congruence of non-expanding, non-shearing, and non-twisting
null geodesics $\bl$ are summarized in table \ref{tab:xKS:Kundt}.

\fulltable{\label{tab:xKS:Kundt}Classification of Kundt spacetimes according to
  the geometry of the geodetic non-expanding, non-shearing, and non-twisting null
  vector field $\bl$. For each subclass, the corresponding holonomy group and
  admissible Weyl type are presented. The Ricci rotation coefficients $L_{11}$
  and $L_{1i}$ are frame components of the covariant derivative of $\bl$ in the
  canonical form \eqref{Kundt:nabla_l}. In the case $L_{1i} \neq 0$, it is always
  possible to set $L_{11} = 0$.
}
  \br
  $L_{11}$ & $L_{1i}$ & Class & Holonomy & Ricci tensor & Weyl type\\
  \mr
  any value & non-zero & Kundt     & $SO(1, n - 1)$ & $R_{00} = 0$                   & I(b) \\
   non-zero &        0 & RNV       & $Sim(n - 2)$   & $R_{00} = R_{0i} = 0$          & II(d) \\
          0 &        0 & \pp waves & $E(n - 2)$     & $R_{00} = R_{0i} = R_{01} = 0$ & II(d) \\
  \br
\endfulltable